\documentclass[twoside,twocolumn,9pt]{article}
\usepackage[super,sort&compress,comma]{natbib} 
\usepackage[left=1.5cm, right=1.5cm, top=1.785cm, bottom=2.0cm]{geometry}
\usepackage{times,mathptmx}
\usepackage{graphicx} 
\usepackage[format=plain,justification=raggedright,singlelinecheck=false,font={stretch=1.125,small,sf},labelfont=bf,labelsep=space]{caption}
\usepackage{float}
\usepackage[english]{babel}
\usepackage{array}
\usepackage[T1]{fontenc}
\usepackage[usenames,dvipsnames]{xcolor}
\usepackage{setspace}
\usepackage[compact]{titlesec}

\usepackage{amssymb,latexsym,amsmath}
\usepackage{color}
\usepackage{ulem}
\usepackage{bm}
\usepackage[babel=true]{csquotes}
\usepackage{subfig}
\usepackage{subeqnarray}
\usepackage{textcomp}
\usepackage{slashbox}
\usepackage[capitalize]{cleveref}

\newcommand*\diff{\mathop{}\!\mathrm{d}}

\usepackage{epstopdf}

\begin{document}

\twocolumn[
  \begin{@twocolumnfalse}
\sffamily
\begin{tabular}{m{4.5cm} p{13.5cm} }

 & \noindent\LARGE{\textbf{Enhanced steady-state dissolution flux in reactive convective dissolution}} \\
\vspace{0.3cm} & \vspace{0.3cm} \\

~ & \noindent\large{V. Loodts\textit{$^{a}$}, B. Knaepen\textit{$^{b}$}, L. Rongy\textit{$^{a}$} and A. De Wit$^{\ast}$\textit{$^{a}$}} \\

\vspace{0.3cm} 

 ~ & \noindent\normalsize{Chemical reactions can accelerate, slow down or even be at the very origin of the development of dissolution-driven convection in partially miscible stratifications, when they impact the density profile in the host fluid phase. 
We numerically analyze the dynamics of this reactive convective dissolution in the fully developed non-linear regime  for a phase A dissolving into a host layer containing a dissolved reactant B.
We show that for a general A+B$\rightarrow$C reaction in solution, the dynamics vary with the Rayleigh numbers of the chemical species, i.e. with the nature of the chemicals in the host phase. 
Depending on whether the reaction slows down, accelerates or is at the origin of the development of convection, the spatial distributions of species A, B or C, the dissolution flux and the reaction rate are different. 
We show that chemical reactions enhance the steady-state flux as they consume A and can induce more intense convection than in the absence of reactions.
This result is important in the context of CO$_2$ geological sequestration where quantifying the storage rate of CO$_2$ dissolving into the host oil or aqueous phase is crucial to assess the efficiency and the safety of the project. 
} 

\end{tabular}

 \end{@twocolumnfalse} \vspace{0.6cm}

  ]

\renewcommand*\rmdefault{bch}\normalfont\upshape
\rmfamily
\section*{}
\vspace{-1cm}

\footnotetext{\textit{$^{\ast}$~Corresponding author; E-mail: adewit@ulb.ac.be}}
\footnotetext{\textit{$^{a}$~Universit\'e libre de Bruxelles (ULB), Facult\'e des Sciences, Nonlinear Physical Chemistry Unit, CP231, 1050 Brussels, Belgium.}}
\footnotetext{\textit{$^{b}$~Universit\'e libre de Bruxelles (ULB), Facult\'e des Sciences, Fluid and Plasmas Dynamics Unit, CP231, 1050 Brussels, Belgium.}}




\section{Introduction} 
Dissolution-driven convection can develop in partially miscible stratifications when the dissolution of a phase A with a finite solubility into a host fluid phase creates an unstable density stratification. 
This can happen for instance when the phase A, dissolving from above, increases the density of the host phase, thereby forming a layer of denser fluid rich in A on top of less dense fluid.
Studying such convective dissolution can help improve the safety of nuclear reactors \cite{nuclear_convection}, optimize industrial production of chemicals \cite{wylock14,wylock16} or understand the physicochemical processes at hand during CO$_2$ geological sequestration \cite{huppert14,emami15}.
The temporal evolution of the dissolution-driven convective dynamics has been characterized in detail and it has been shown that the dissolution flux reaches a steady-state value before shutdown \cite{pau10,elenius12,slim13,slim14,emami15}.
Understanding the impact of chemical reactions on such dynamics has recently gained interest because of the potential effect of geochemistry on the efficiency of CO$_2$ geological sequestration \cite{emami15,Geochemistry}.
Reactions can indeed affect convection because they modify solute concentrations affecting fluid properties such as the density of the solution, leading to different possible scenarios for the development of convection \cite{dewit16,loodts16}.

When the dissolving species A reacts with a solute B initially present in the host solution following a second-order scheme A+B $\rightarrow$ C, the reaction can slow down or accelerate the development of convection compared to the non-reactive case depending on the type of density profile building up in the host phase, as shown by both experimental and numerical approaches \cite{loodts14prl,loodts15,budroni14,budroni17,cherezov16,kim16,wylock14,wylock16}. 
If C is less dense than B, a density profile with a minimum is observed and the instability develops more slowly than in the non-reactive case.
On the contrary, if C is sufficiently denser than B, the instability develops faster and the density profile is monotonic like its non-reactive counterpart. 
A simpler reaction where B is solid and no C is produced reduces the growth rate of convection, because A at the origin of the instability is consumed by the reaction \cite{ ennis07, ghesmat11, andres11, andres12, cardoso14, kim14, kim15, ward14lsa, ward14nl,ward15}.
In contrast, when A decreases the density of the solution upon dissolution, the non-reactive case is buoyantly stable and reactions can be at the origin of a density profile unstable with regard to convection due to the creation of a maximum of density \cite{loodts15,kim16}.

The different possible convective dynamics in the presence of reaction have been classified according to whether the reaction slows down or accelerates the development of dissolution-driven convection \cite{loodts14prl,loodts15,kim16,budroni17}.
We here broaden the scope of this classification by analyzing the effects of reaction on other aspects of the convective dynamics in the fully developed non-linear regime. 
First, it is crucial to understand whether the different successive regimes of the convective dynamics identified in the non-reactive case can also be observed in reactive cases. 
Second, for potential applications it is of interest to characterize the impact of reactions on the evolution of the flux of A dissolving into the host phase, in particular on its steady-state value already quantified in the non-reactive case \cite{pau10, elenius12, hidalgo12, slim13, slim14, elenius15, islam15}, and on the global reaction rate in the host phase. 
Finally, we aim to characterize the effects of convection on the dynamics of the reaction front, which also remain poorly understood. 

We address these issues by theoretically studying the effects of an A+B$\rightarrow$C reaction on the non-linear dynamics during dissolution-driven convection in a host fluid phase occupying a porous medium, relevant to the context of CO$_2$ sequestration. 
Similarly to previous studies \cite{loodts14prl,loodts15,budroni14,ghoshal17,cherezov16}, we consider equal diffusivities of the three chemical species in order to focus solely on solutal effects. 
By varying the relevant control parameters, we study their effect on the properties of the reaction-diffusion-convection dynamics: the evolution of the fingering pattern, the dynamics of the reaction front, the dissolution flux and the global reaction rate affecting the storage of A into the host fluid are all quantified. 
We qualitatively describe the dynamics observed in the unstable non-reactive case and in three typical reactive cases: less unstable or more unstable than the unstable non-reactive counterpart, as well as destabilization by reaction of a stable non-reactive counterpart.
Further, we derive simple scalings to predict the steady-state reaction rate and dissolution flux of A at the interface as a function of the Rayleigh numbers.
Such scalings could be used to predict the temporal evolution of the quantity of A in the host phase.
 
\section{Model}\label{model}
\begin{figure}[htbp]
\centering
\includegraphics[width=\columnwidth]{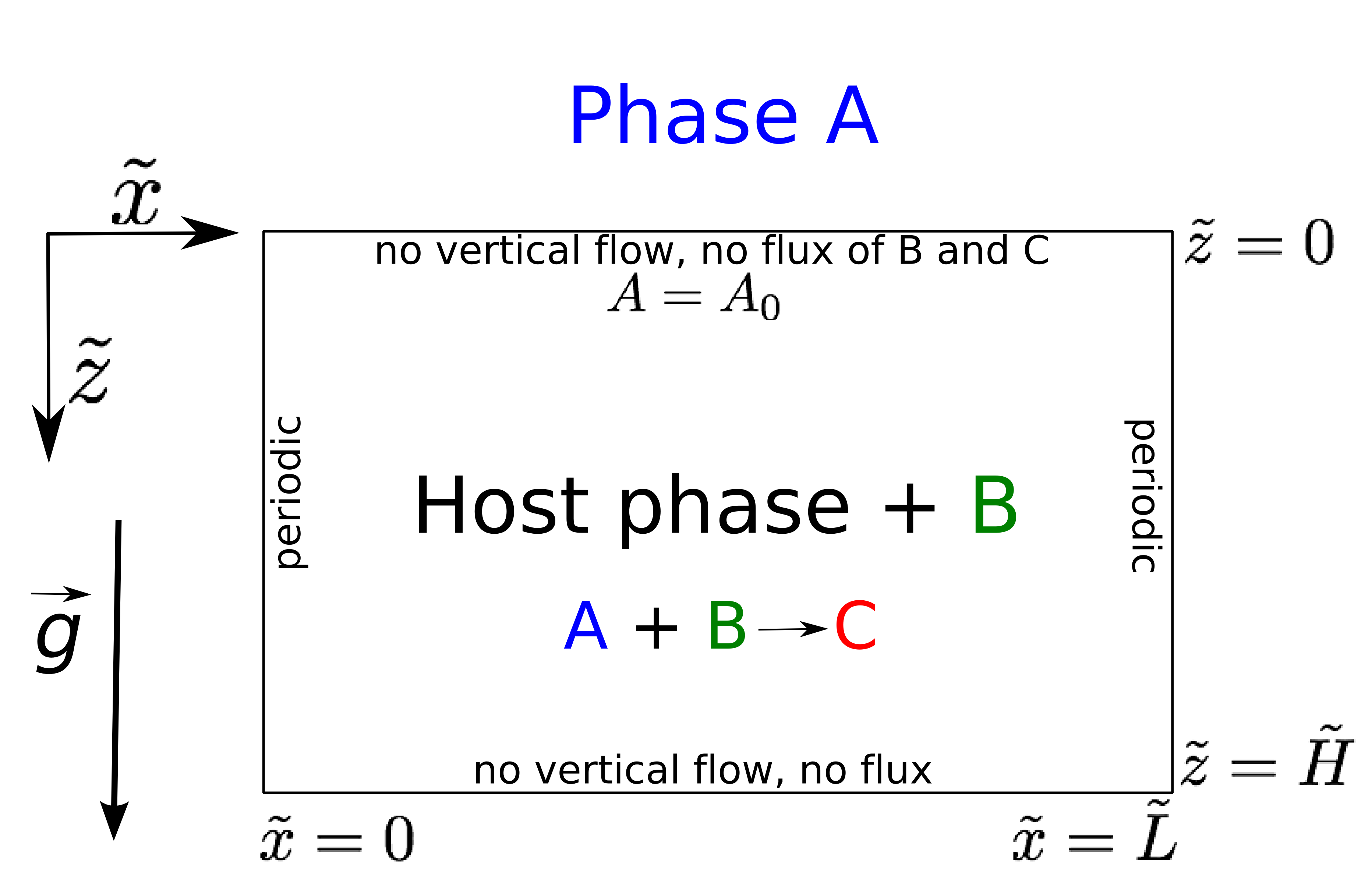}
\caption{Schematic of the bidimensional system.}
\label{setup}
\end{figure}

We consider an isothermal, isotropic and homogeneous vertical system, in which two partially miscible phases are placed in contact along a horizontal interface in a porous medium \cite{loodts14prl, loodts15}. 
The gravity field $\bm g$ points downwards, along the vertical $\tilde z$ axis perpendicular to the horizontal $\tilde x$ axis.
Phase A dissolves into the other lower fluid phase, called ``host'' phase as shown in Fig. \ref{setup}.
The concentration of A at the interface is considered to remain constant over time, and to be equal to its solubility $A_0$ in the host phase, following the assumption of local chemical equilibrium. 

The host phase contains a reactant B dissolved with an initial concentration $B_0$.
Species A reacts with solute B to produce another solute C, following a second-order A+B$\rightarrow$C reaction.
All three species can thus contribute to changes in density.
The concentrations of B and C are assumed small enough and therefore do not significantly affect the solubility $A_0$ of A into the host phase.

The interface is considered to be permeable to species A but impermeable to the solvent of the host phase and to solutes B and C. 
To focus on the effects of the reaction on the dynamics, we assume that the interface remains in the course of time at the same position $\tilde z=0$ and we study the dynamics in the host phase only. 
The host phase extends from $\tilde x = 0$ to $\tilde L$ in the horizontal direction and from the interface at $\tilde z=0$ to $\tilde z = \tilde H$ in the vertical direction.

An equation for the fluid flow velocity $\bm{\tilde u}=(\tilde u_x, \tilde u_z)$ of an incompressible flow is coupled to the reaction-diffusion-convection (RDC) equations for solute concentrations via an equation of state for the density depending linearly on the concentrations\cite{loodts14prl,loodts15}.
The concentrations, time, spatial coordinates and velocity are normalized using \cite{loodts14prl,loodts15}:
	\begin{eqnarray}\label{eq: variables} 	
	 A &=  \tilde A/A_0, \quad B =  \tilde B/A_0, \quad C =  \tilde C/A_0,   \label{eq: variables concentration}\\
	 t &= \tilde t /t_c , \quad  \bm {z} = \bm { \tilde  z} /l_c,  \quad \bm {u} = \bm { \tilde u}/u_c, \label{eq: variables u l t} 
\end{eqnarray}
where tildes denote dimensional variables.
We nondimensionalize the solute concentrations $\tilde A$, $\tilde B$ and $\tilde C$ in Eq. \eqref{eq: variables concentration} with the solubility $A_0$ of phase A. 
In Eq. \eqref{eq: variables u l t}, we use the chemical time scale $t_c=1/(q A_0)$ with $q$ the kinetic constant of the reaction A+B$\rightarrow$C, the RD length scale $l_c=\sqrt{D_A t_c}=\sqrt{D_A/(q A_0)}$ with $D_A$ the diffusion coefficient of A and the velocity scale $u_c =\phi l_c/t_c = \phi \sqrt{D_A q A_0}$ with $\phi$ the porosity of the medium at hand.

The dimensionless RDC equations then read:
	\begin{eqnarray}\label{eq: rdc1}
	\frac{\partial A}{\partial t}+(\bm u \cdot \boldsymbol \nabla) A  &=& \nabla^2 A -A B, \label{eq: a1}\\
	\frac{\partial B}{\partial t}+(\bm u \cdot \boldsymbol \nabla) B  &=& \delta_B \nabla^2 B - A B, \label{eq: b1}\\
	\frac{\partial C}{\partial t}+(\bm u \cdot \boldsymbol \nabla) C  &= &\delta_C \nabla^2 C + A B, \label{eq: c1}
	\end{eqnarray}
where $\delta_B=D_B/D_A$ and $\delta_C = D_C/D_A$, with $D_B$ and $D_C$ the diffusion coefficients of species B and C. 
Using the notations $H=\tilde H/l_c$ and $L = \tilde L/l_c$ for the dimensionless height and width of the host phase, we solve Eqs. \eqref{eq: a1}-\eqref{eq: c1} with the boundary conditions (see Fig. \ref{setup})
	\begin{eqnarray}\label{cb}
	u_z(z=0) = 0  ;& \quad u_z(z=H) = 0 ; \quad \bm{u}(x=0) = \bm{u}(x=L), \label{cb u} \\
	 A(z=0) = 1  ;& \quad \left.\frac{\partial A}{\partial  z}\right\rvert_{ z=H} = 0 ; \quad A(x=0) = A(x=L),\label{cb a}\\
	\left.\frac{\partial B}{\partial  z}\right\rvert_{z=0} = 0 ;& \quad  \left.\frac{\partial B}{\partial  z}\right\rvert_{ z=H}= 0 ; \quad B(x=0) = B(x=L), \label{cb b} \\
	\left.\frac{\partial C}{\partial z}\right\rvert_{z=0} = 0 ;& \quad  \left.\frac{\partial C}{\partial  z}\right\rvert_{z=H} = 0 ; \quad C(x=0) = C(x=L), \label{cb c} 
	\end{eqnarray}
and the initial conditions
	\begin{eqnarray}\label{ci}
	A(x,z=0,t=0)=1+\epsilon \cdot {\rm rand}(x); A(x,z>0,t=0)=0, \label{ci a} \\
	B(x,z,t=0)  = \beta=B_0/A_0, \label{ci b} \\
	C(x,z,t=0)  = 0, \label{ci c}
	\end{eqnarray}
where $\beta = B_0/A_0$ is the ratio between the initial concentration $B_0$ of reactant B and the solubility $A_0$ of A in the host phase. 
When $\beta = 0$, the non-reactive case is recovered; when $\beta \to \infty$, reactant B is in large excess with regard to A so that the reaction can be considered first-order at early times \cite{andres11, andres12, cardoso14, kim14, kim15, ward14lsa, ward14nl, ghoshal17}.
Equation \eqref{ci a} expresses that perturbations are introduced in the initial concentration of A at the interface in order to trigger the instability (see e.g. Refs. \citenum{bestehorn12,tilton13} for a discussion of the possible types of perturbations).
$\epsilon \ll 1$ is the amplitude of the perturbation, here chosen as $10^{-3}$, and rand$(x)$ is its modulation,  function of the horizontal coordinate $x$ and varying randomly between -1 and 1 (``white noise'').

The set of equations \eqref{eq: a1}-\eqref{eq: c1} are closed using an equation for the fluid flow velocity of an incompressible flow. 
To that end, we assume a linear state equation for the dimensional density $\tilde \rho$ of the solution as a function of the solute concentrations:
	\begin{equation}
	\tilde \rho = \rho_0 (1+\alpha_A \tilde A +\alpha_B \tilde B +\alpha_C \tilde C),
	\end{equation}
where $\alpha_i = \frac{1}{\rho_0}\frac{\partial \rho}{\partial c_i}$ is the solutal expansion coefficient of species i. 	
A dimensionless density can be computed as
	\begin{equation}
	\frac{\tilde \rho - \rho_0}{\rho_c} = R_A  A+R_B  B +R_C  C, \label{eq: density1} 
	\end{equation}
where $\rho_c = \phi \mu D/(g \kappa   l_c)$ is the density scale and the Rayleigh numbers $R_i$ (i=A,B,C) quantify the contribution of species i to the dimensionless density of the solution, constructed with the RD length scale (Eq. \eqref{eq: variables u l t}):
	\begin{equation} \label{eq: rayleigh}
	R_i = \frac{\alpha_i A_0 g \kappa   l_c}{\phi \nu D_A} 
	= \frac{\alpha_i A_0 g \kappa}{\phi \nu \sqrt{D_A q A_0}},
	\end{equation}
with $\nu = \mu/\rho_0$ the kinematic viscosity of the solvent. 
The expression \eqref{eq: density1} is appropriate in the general case where the species have different diffusivities. 
For $\delta_i \neq 1$, the problem is thus dependent on six parameters: $\delta_B$, $\delta_C$, $R_A$, $R_B$, $R_C$ and $\beta$. 

We further assume here that all species A, B and C have the same diffusion coefficient so that $\delta_B = \delta_C =1$.
We can then add Eqs. \eqref{eq: b1} and \eqref{eq: c1}, taking into account the corresponding boundary (Eqs. \eqref{cb b}-\eqref{cb c}) and initial (Eqs. \eqref{ci b}-\eqref{ci c}) conditions, to obtain the conservation relation
\begin{equation}\label{c computed}
 B = \beta - C.
\end{equation}
Thanks to this conservation relation, we can reduce the number of dimensionless parameters further by defining the dimensionless density $\rho$ as
	\begin{equation}\label{eq: density_2}
	\rho = \frac{\tilde \rho - \rho_0}{\rho_c}  - R_B \beta,
	\end{equation}
i.e.
	\begin{equation}
	\rho = R_A  A+ \Delta R_{CB} C,\label{eq: rho3}
	\end{equation}
where $\Delta R_{CB} = R_C - R_B$ represents the difference between the contributions to density of product C and reactant B.
For equal diffusivities $\delta_B=\delta_C=1$, the definition \labelcref{eq: rho3} of $\rho$ explicitly highlights that the system is characterized by only three parameters, here chosen as\footnote{Note that our previous results\cite{loodts14prl,loodts15} remain the same with this new formulation \labelcref{eq: rho3} for $\rho$ because they were performed for $R_B=0$, so that the parameter $R_C$ that we varied is strictly equivalent to $\Delta R_{CB}$ here.} $R_A$, $\Delta R_{CB}$ and $\beta$. 
Darcy's equation and the incompressibility condition expressed as Poisson equation then read in dimensionless form:
\begin{eqnarray}\label{flow}
\boldsymbol{\nabla}  p  &=  -\bm{ u}+ \rho \bm{e_z}, \label{eq: darcy} \\
	{\nabla}^2  p  &=  \boldsymbol{\nabla} \cdot (\rho \, \bm {e_z}), \label{eq: poisson}
\end{eqnarray}
with $p$ the dimensionless pressure.

We numerically solve Eqs. \eqref{eq: a1}-\eqref{eq: c1} with $\delta_B=\delta_C=1$, Eqs. \eqref{eq: rho3}-\eqref{eq: poisson} with Eqs. \eqref{cb u}-\eqref{ci c} on a computational domain of width $L=3072$ and height $H=2048$ using the \texttt{YALES2} software \cite{moureau11}, more specifically the \texttt{DARCY\_SOLVER} module \cite{loodts16phd}. 
This software is based on the finite volume method \cite{CFD}.
We use an explicit method called TFV4A or TRK4 (two-step Runge-Kutta with a fourth-order spatial discretization) \cite{kraushaar11}.
The dynamics depend on the random noise added to the initial condition in Eq. \eqref{ci a}. 
Therefore, for each value of the set of parameters ($R_A,\Delta R_{CB},\beta$), we average the results over 15 realizations to obtain robust results. 
Increasing that number of realizations above 15 does not impact the averages and standard deviations of the results significantly (below 5\%). 
The uncertainty linked to the different possible noises is quantified as the 95 \% confidence interval for two-sided critical regions.
In addition, we have checked that these results averaged over 15 realizations were robust with regard to refinement of the iterative convergence tolerance (here 10$^{-10}$) for solving Poisson's equation \eqref{eq: poisson} with \texttt{HYPRE}, mesh size ($\Delta x=\Delta z=4$) and time step ($\Delta t=0.5$).
With these values, the iterative convergence errors and discretization errors on the results were smaller than 5\%. 

We perform a parametric study of the non-linear dynamics as a function of $R_A$ and $\Delta R_{CB}$ while keeping $\beta=1$. 
Indeed it is already known that increasing $\beta$, i.e. amplifying the amount of dissolved reactant B with regard to the solubility of A, amplifies the effect of reaction on the development of dissolution-driven convection. 
If the instability develops faster than in the non-reactive case, a larger $\beta$ accelerates even more that development; conversely if chemistry slows down the growth of convection, increasing $\beta$ decreases the growth rate of the instability \cite{loodts14prl, loodts15, budroni14, budroni17, cherezov16}.
Our objective here is to analyze the effect of changing the nature of reactants on dissolution-driven convection. 
We therefore analyze two main classes of dissolving species A, taking $R_A$=+1 (-1) representing a component increasing (respectively decreasing) the density of the host phase upon dissolution. 
Fixing thus $|R_A|=1$ and only varying the sign of $R_A$ affects the type of density profile above the reaction front \cite{loodts16, loodts15}: increasing downwards if $R_A=-1$ (stable non-reactive counterpart) or decreasing downwards if $R_A=1$ (unstable non-reactive counterpart).
For each case, we vary $\Delta R_{CB}$, which corresponds to scanning various possible reactant B and product C pairs. 
To study the effect of the composition and, in particular, of solutal effects on the convective dynamics, we analyze the fingering and reaction zone evolution, the dissolution flux, and the volume-averaged concentrations as a function of $\Delta R_{CB}$ varying between -1 and +1.  
 
\section{Dissolution-driven convective dynamics}\label{fingering dynamics}
We start by qualitatively describing the dissolution-driven convective dynamics in a few specific cases without reaction (Section \ref{nr_pictures}) or with reaction (Section \ref{r_pictures}).
We then compare the fingering dynamics using space-time plots (Section \ref{space-time sec}). 

\subsection{Non-reactive cases}\label{nr_pictures}
The non-reactive case can be either unstable ($R_A>0$) or stable ($R_A<0$) with regard to dissolution-driven convection. 
\subsubsection{NR case: non-reactive unstable ($R_A=1$)}
\begin{figure}[bt]\begin{center}
\subfloat[][$t=2000$]{\includegraphics[width=0.49\columnwidth]{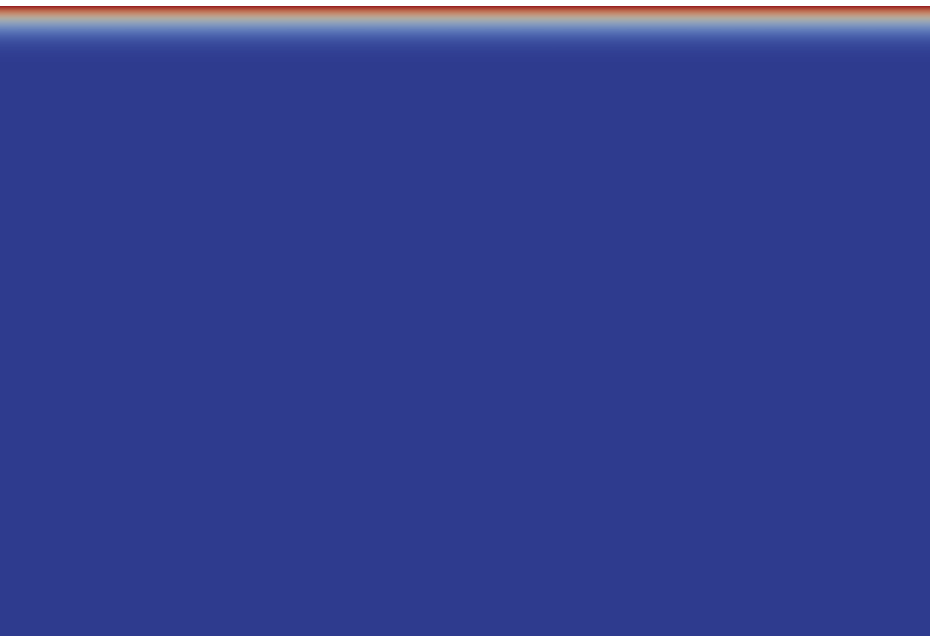}\label{density field nr25}}\hfill
\subfloat[][$t=4000$]{\includegraphics[width=0.49\columnwidth]{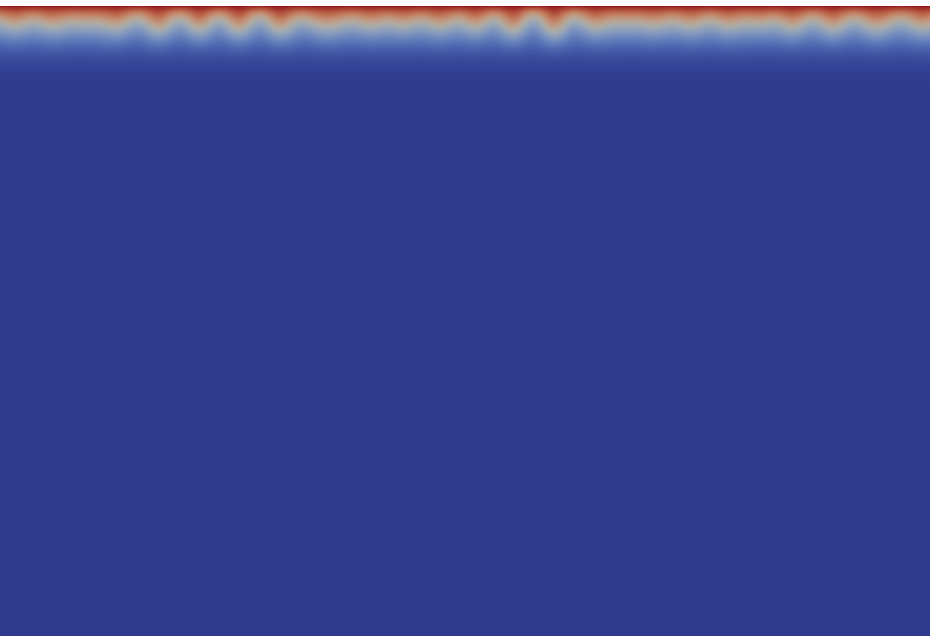}\label{density field nr50}}\\[-2.5ex]
\subfloat[][$t=8000$]{\includegraphics[width=0.49\columnwidth]{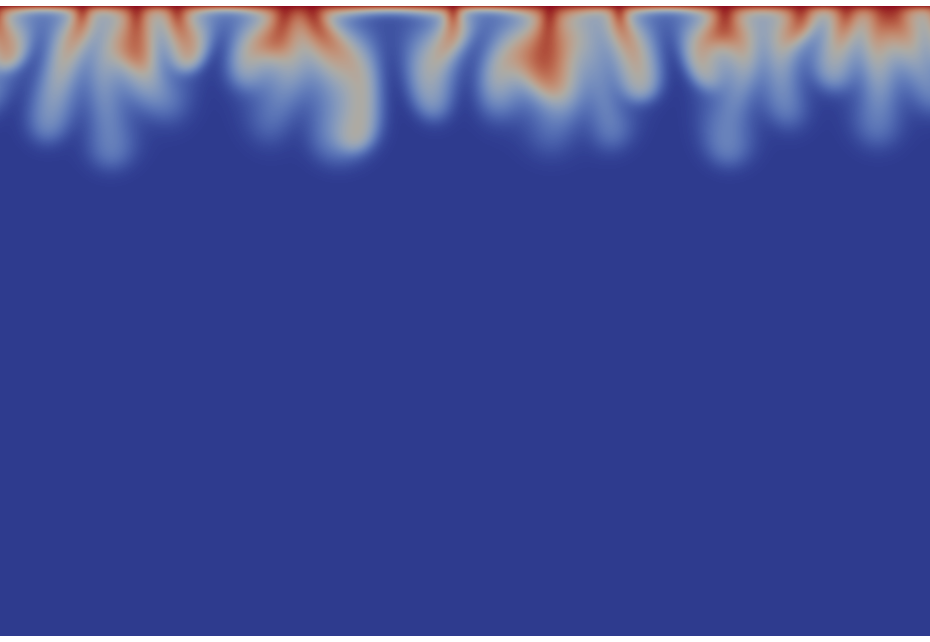}\label{density field nr75}}\hfill
\subfloat[][$t=12000$]{\includegraphics[width=0.49\columnwidth]{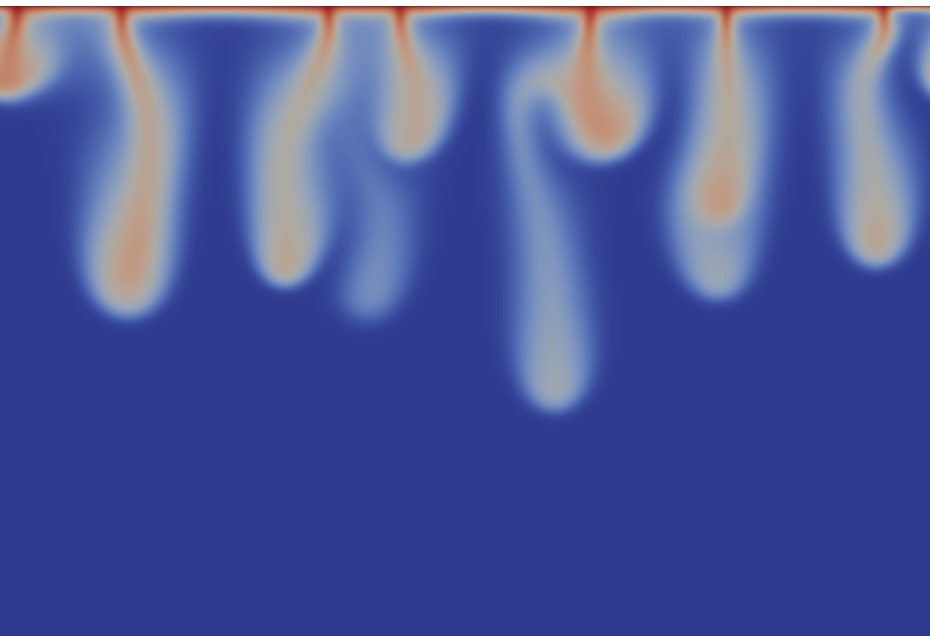}\label{density field nr100}}\\[-2.5ex]
\subfloat[][$t=16000$]{\includegraphics[width=0.49\columnwidth]{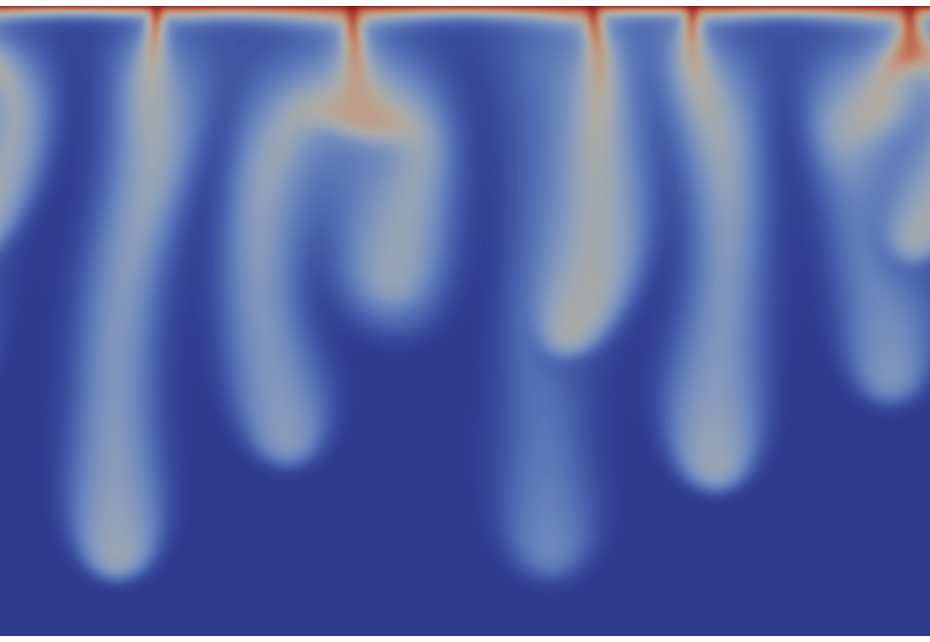}\label{density field nr150}}\hfill
\subfloat[][$t=24000$]{\includegraphics[width=0.49\columnwidth]{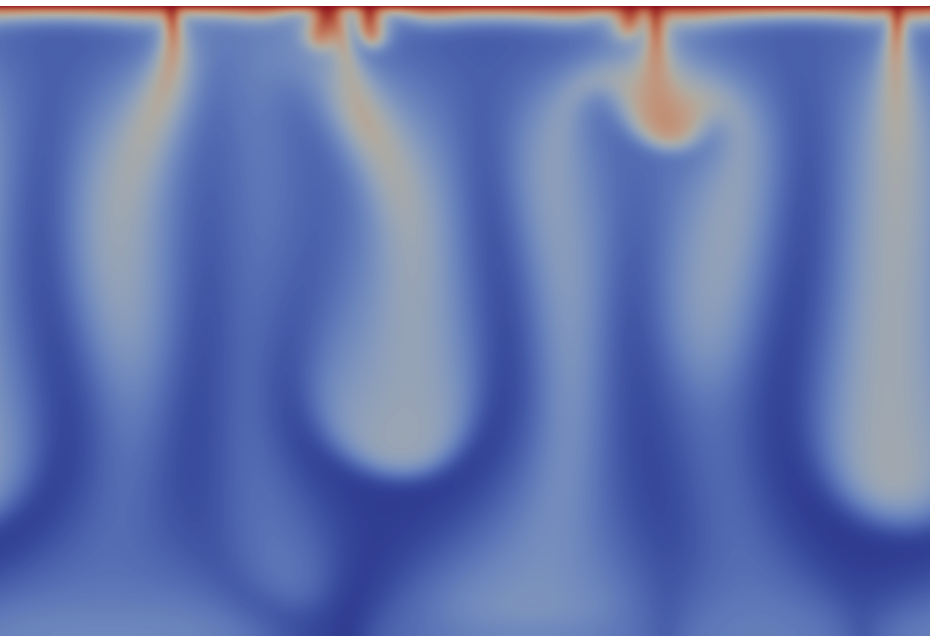}\label{density field nr200}}
\caption{Density field in the host solution of dimensions 3072$\times$2048 in the unstable non-reactive case (NR,  $R_A=1$) at different times $t$ for a typical realization. 
The density scale varies between 0 (blue) and 1 (red).}
\label{density field nr}
\end{center}\end{figure}

The non-reactive (NR) unstable case for which $R_A=1$ has been well characterized in the literature \cite{pau10,elenius12,slim13,slim14,emami15}.
If species A increases the density of the host phase ($R_A > 0$), the dissolution of A into a lower host phase progressively creates a buoyantly unstable density stratification.
The dynamics in this unstable non-reactive case are illustrated for a specific realization in Fig. \ref{density field nr} showing the density field at different times.
The dynamics can be divided in different successive regimes \cite{slim13, slim14}.
Initially, the miscible contact zone between the denser zone rich in A below the interface and the less dense bulk solution below it is flat (Fig. \ref{density field nr25}). 
This zone deforms gradually once fingers of the denser fluid begin to sink into the lower part of the host phase (Fig. \ref{density field nr50}).
At the beginning these fingers do not interact significantly with each other and a well defined wavelength is observed: soon after the onset of the instability, 22 fingers can be observed on the total width of 3072 (see Fig. \ref{density field nr50}), which corresponds to a wavelength of $\approx$ 140, in agreement with the results of other non-linear simulations \cite{slim14,elenius12}.
After some time, merging becomes a dominant process and the number of fingers decreases dramatically (Figs. \ref{density field nr75}-\ref{density field nr100}). 
After this merging regime, the number of fingers only decreases slightly (Figs. \ref{density field nr150}-\ref{density field nr200}).
In the reinitiation regime \cite{hewitt13, slim14}, small new fingers, called protoplumes, develop from the boundary layer and then join older fingers.

\subsubsection{NR2 case: non-reactive stable ($R_A=-1$)}
If the density decreases upon dissolution of A ($R_A \le 0$), the stratification is buoyantly stable in the absence of reactions as the growing boundary layer rich in A is less dense than the host solvent.
Species A then invades the host phase by diffusion only, and no convection develops (not shown here).

\subsection{Reactive cases}\label{r_pictures}
We now analyze the effect of chemical reactions on the two non-reactive cases presented above. 
To do so, we describe three specific reactive cases as shown in \ref{cases nl}.

\begin{table}[htbp]
\begin{center}
\begin{tabular}{|p{0.25\columnwidth}||p{0.25\columnwidth}|p{0.25\columnwidth}|}
\hline
 \backslashbox{$R_A$}{$\Delta R_{CB}$} & -1 & 1  \\
\hline \hline
1 & \textcolor{ForestGreen}{R1} less unstable reactive  & \textcolor{red}{R2} more unstable reactive\\
\hline
-1 & (not discussed) stable & \textcolor{red}{R3} unstable due to reaction  \\
\hline
\end{tabular}
\end{center}
\caption{Specific reactive cases discussed in \ref{fingering dynamics}: R1, R2 and R3. 
In all cases $\beta = 1$.}
\label{cases nl}
\end{table}%

\subsubsection{R1 case: less unstable reactive system ($R_A=1$, $\Delta R_{CB}=-1$)}
\begin{figure}[htbp]\centering
\subfloat[column 1][$t=4000$]{\includegraphics[width=0.49\columnwidth]{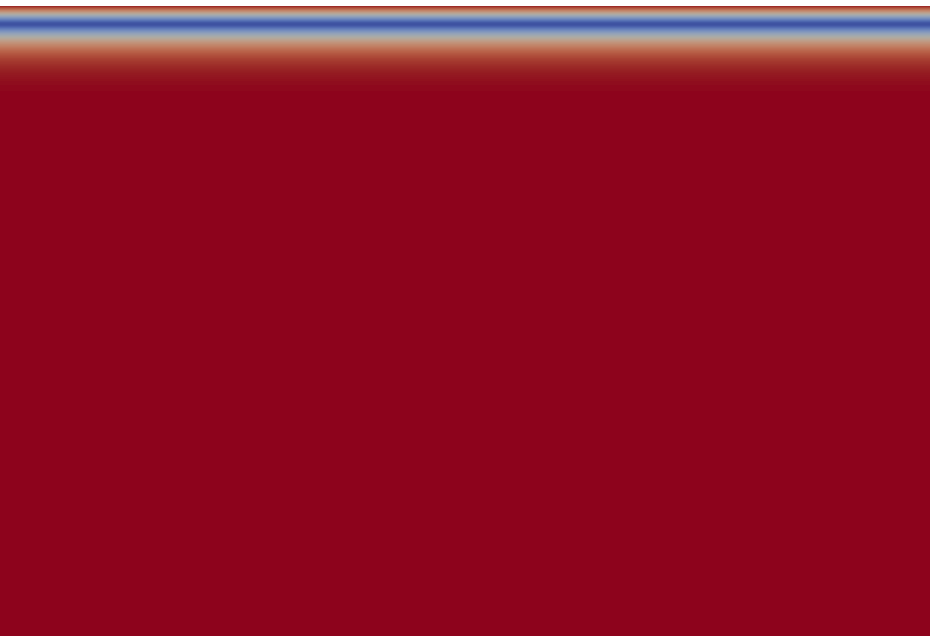}\label{densityField_rc-1_50}}\hfill
\subfloat[column 2][$t=8000$]{\includegraphics[width=0.49\columnwidth]{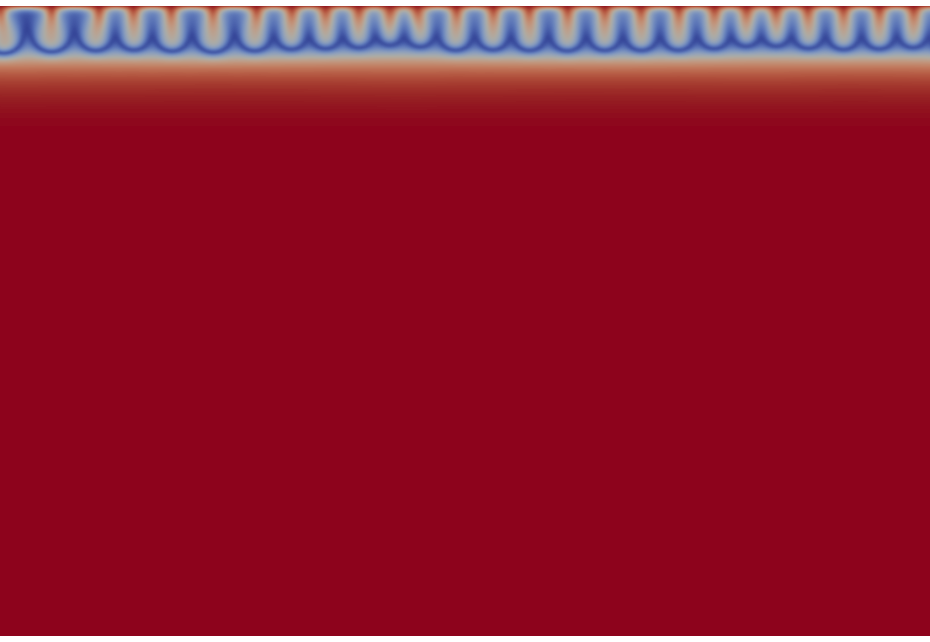}\label{densityField_rc-1_60}}\\[-2.5ex]
\subfloat[column 1][$t=12000$]{\includegraphics[width=0.49\columnwidth]{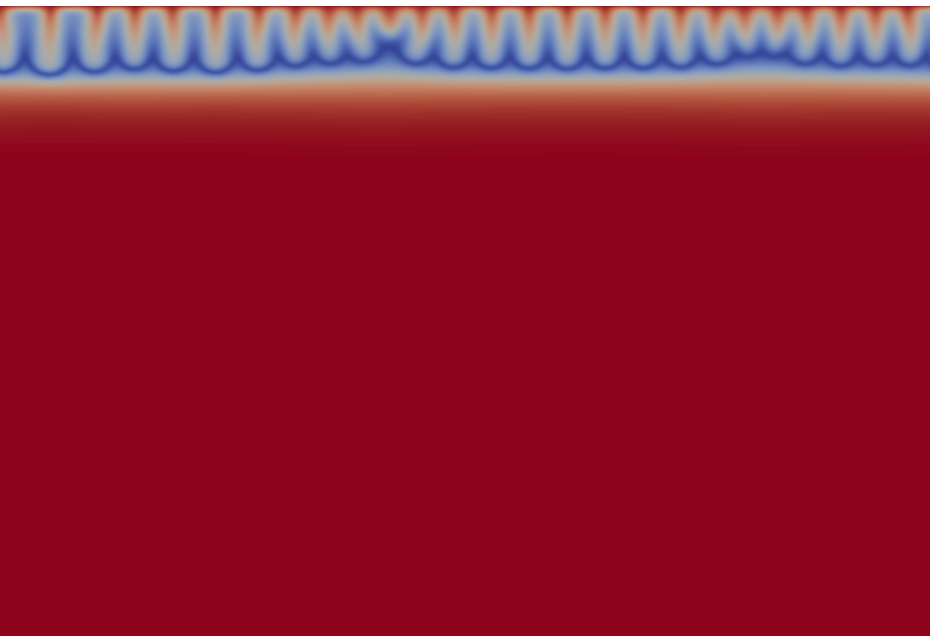}\label{densityField_rc-1_120}}\hfill
\subfloat[column 2][$t=20000$]{\includegraphics[width=0.49\columnwidth]{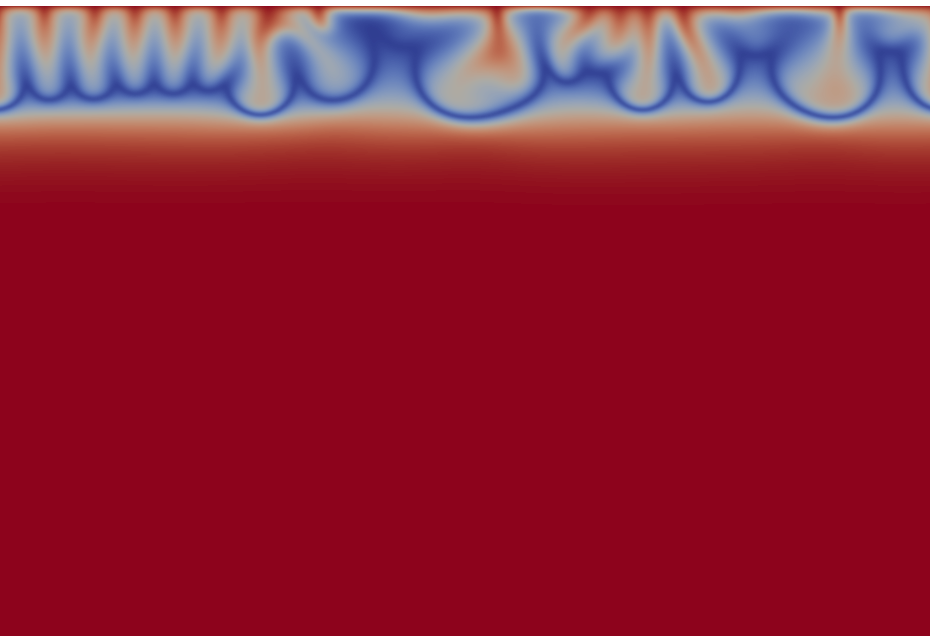}\label{densityField_rc-1_200}}\\[-2.5ex]
\subfloat[column 1][$t=24000$]{\includegraphics[width=0.49\columnwidth]{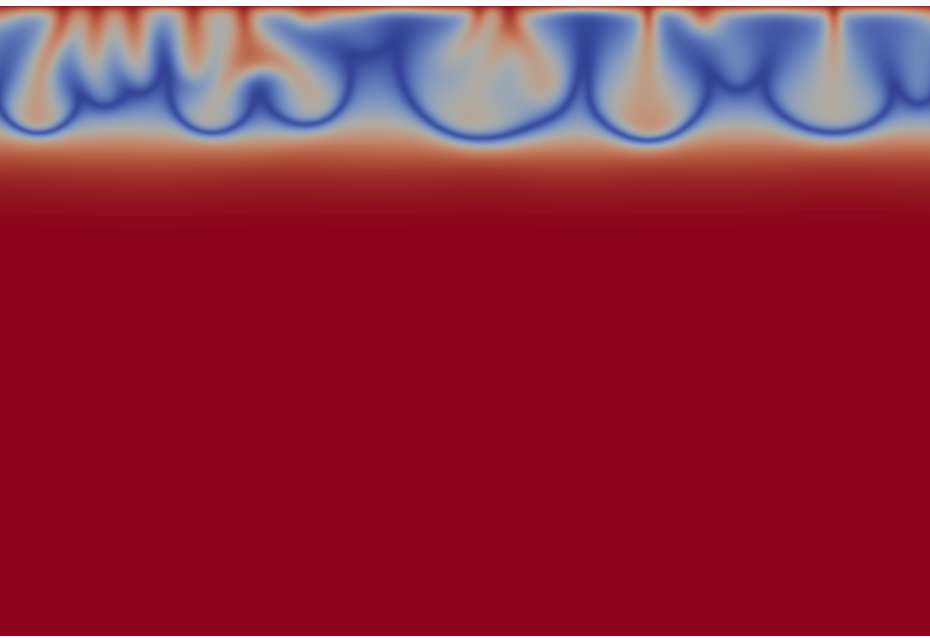}\label{densityField_rc-1_240}}\hfill
\subfloat[column 2][$t=28000$]{\includegraphics[width=0.49\columnwidth]{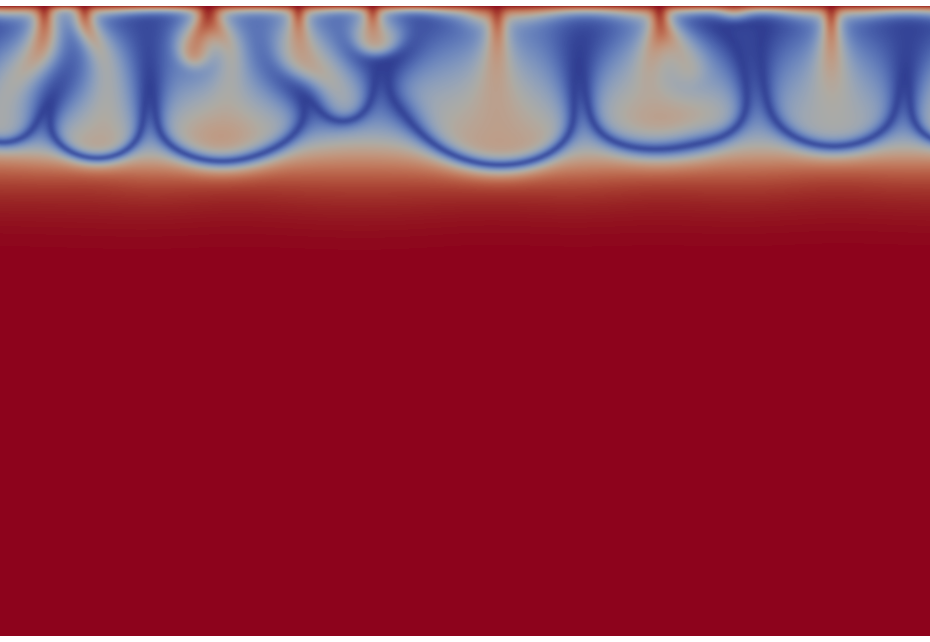}\label{densityField_rc-1_310}}
\caption{
Density field at different times $t$ in R1 case (stabilizing chemistry, $R_A=1$ and $\Delta R_{CB}=-1$). 
The scale varies between -1 (blue) and 0 (red). 
	}
\label{densityField_rc-1}
\end{figure}

When the solution of C is less dense than that of B ($\Delta R_{CB} < 0$), the chemical reaction is expected to slow down the development of fingering because a minimum develops in the density profile at the location of the reaction front \cite{budroni14, budroni17, loodts14prl, loodts15}.
The temporal dynamics of the density field shown in Fig. \ref{densityField_rc-1} indeed illustrate that fingers develop more slowly than in the NR case. 
The fingering dynamics can be described by the same successive regimes as in the NR case: fingers have not developed yet  (Fig. \ref{densityField_rc-1_50}), they grow without interactions between them  (Fig. \ref{densityField_rc-1_60}) and they merge several times (Figs. \ref{densityField_rc-1_120}-\ref{densityField_rc-1_310}).
By contrast to the non-reactive counterpart, protoplumes already form in the merging regime (Fig. \ref{densityField_rc-1_310}). 
The initial number of fingers (see Fig. \ref{densityField_rc-1_60}) is larger than in case NR (Fig. \ref{density field nr50}), i.e. we have here roughly 26 fingers.

\begin{figure}[htbp]\centering
\subfloat[][$A(x,z,t)$]{\includegraphics[width=0.49\columnwidth]{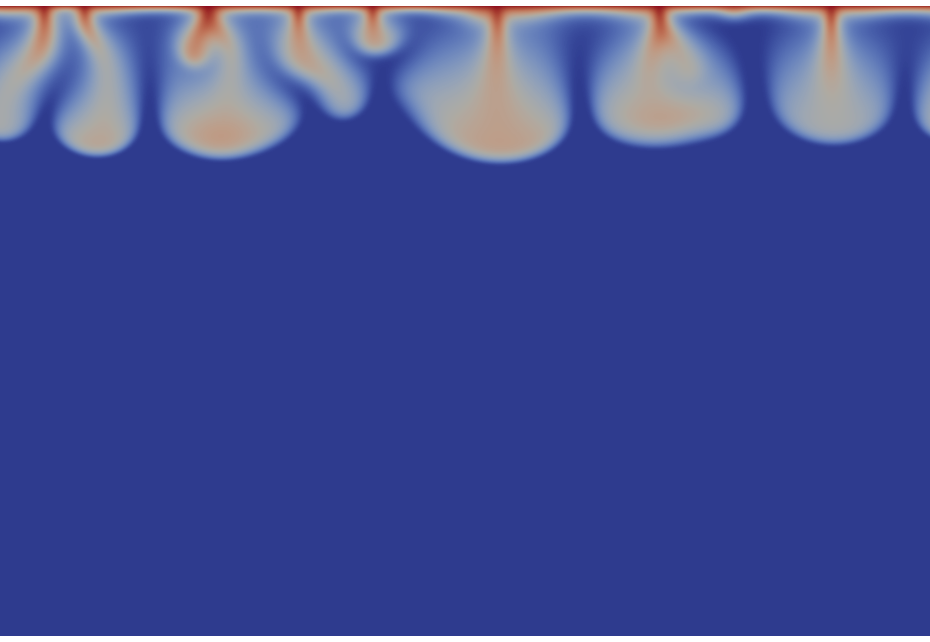}\label{aField_rc-1}}\hfill
\subfloat[][$B(x,z,t)$]{\includegraphics[width=0.49\columnwidth]{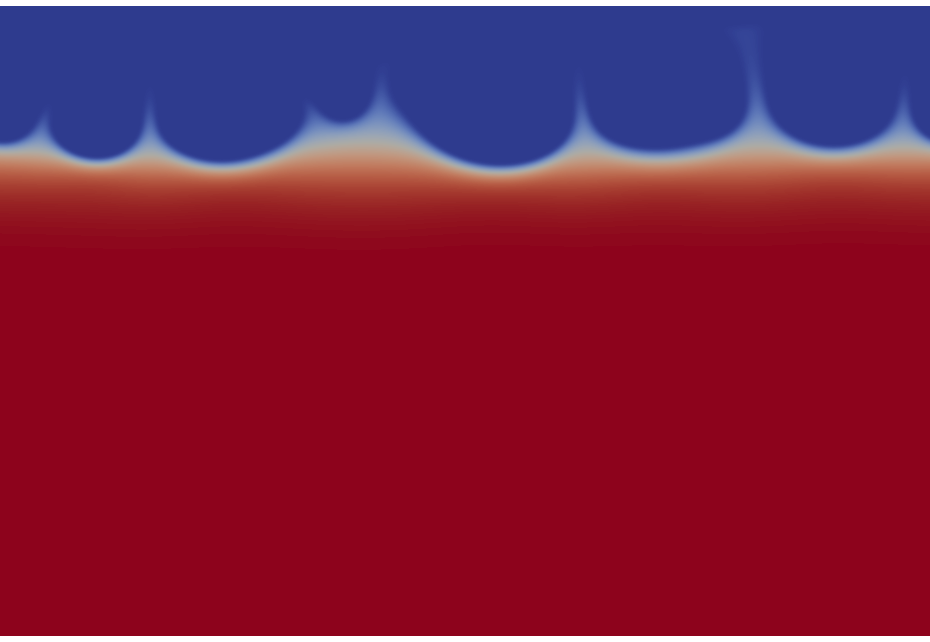}\label{bField_rc-1}}\\[-2.5ex]
\subfloat[][$C(x,z,t)$]{\includegraphics[width=0.49\columnwidth]{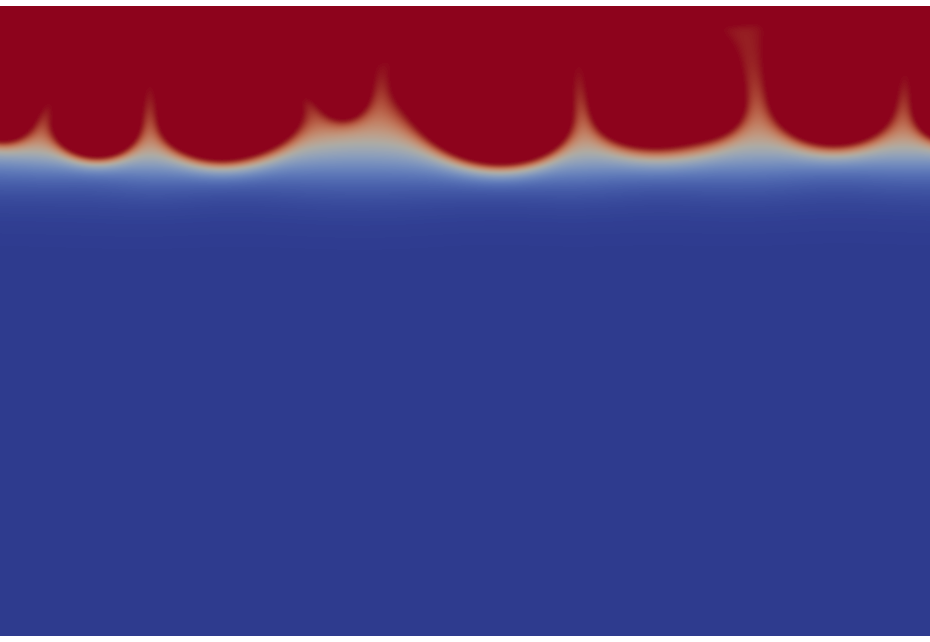}\label{cField_rc-1}} \hfill
\subfloat[][$AB(x,z,t)$]{\includegraphics[width=0.49\columnwidth]{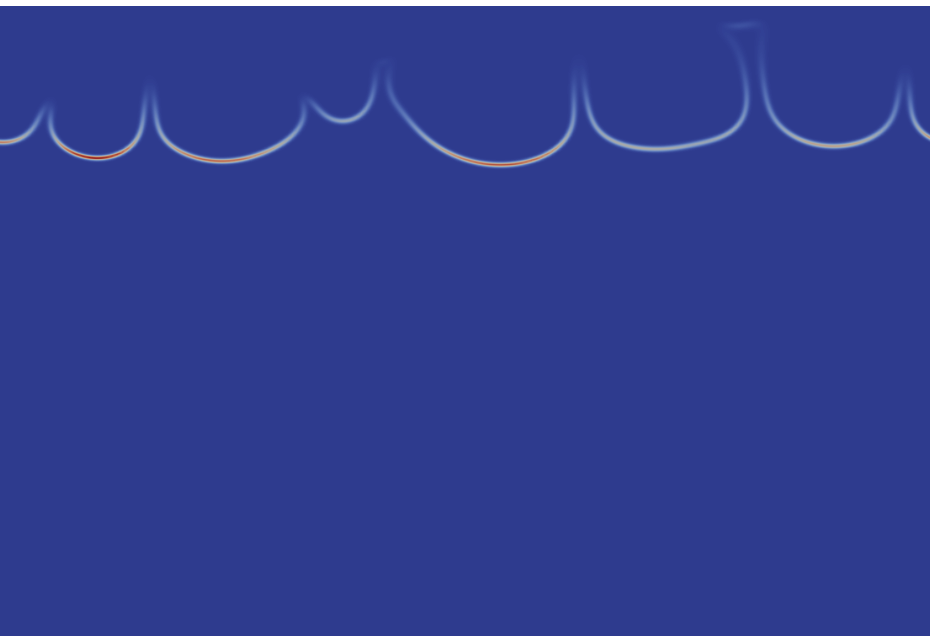} \label{abField_rc-1}}
\caption{Concentration fields of the dissolving species A (a), reactant B (b) and product C (c) varying between 0 (blue) and 1 (red), and reaction rate $AB$ (d) varying between 0 (blue) and 0.002 (red) at time $t=28000$ corresponding to the density field shown in Fig. \ref{densityField_rc-1_310} for case R1.}
\label{concentrationField_rc-1}
\end{figure}

Figures \ref{concentrationField_rc-1} shows the concentration fields responsible for the density field in Fig. \ref{densityField_rc-1_310}. 
The dissolving species A is seen to be at the origin of the fingering pattern (Fig. \ref{aField_rc-1}). 
There are no fingers in the concentration fields of reactant B and product C although the contact line between the reacted (where A and C are present) and unreacted (with mostly B) zones is deformed by the finger tips of A (Figs. \ref{bField_rc-1}-\ref{cField_rc-1}). 
The reaction occurs mostly in a thin zone localized at the tip of the fingers (Fig. \ref{abField_rc-1}), which corresponds to the position of the minimum of density as shown by the comparison of the reaction rate map (Fig. \ref{abField_rc-1}) with the density field (Fig. \ref{densityField_rc-1_310}).

\begin{figure}[tbhp]\centering	
\includegraphics[width=\columnwidth]{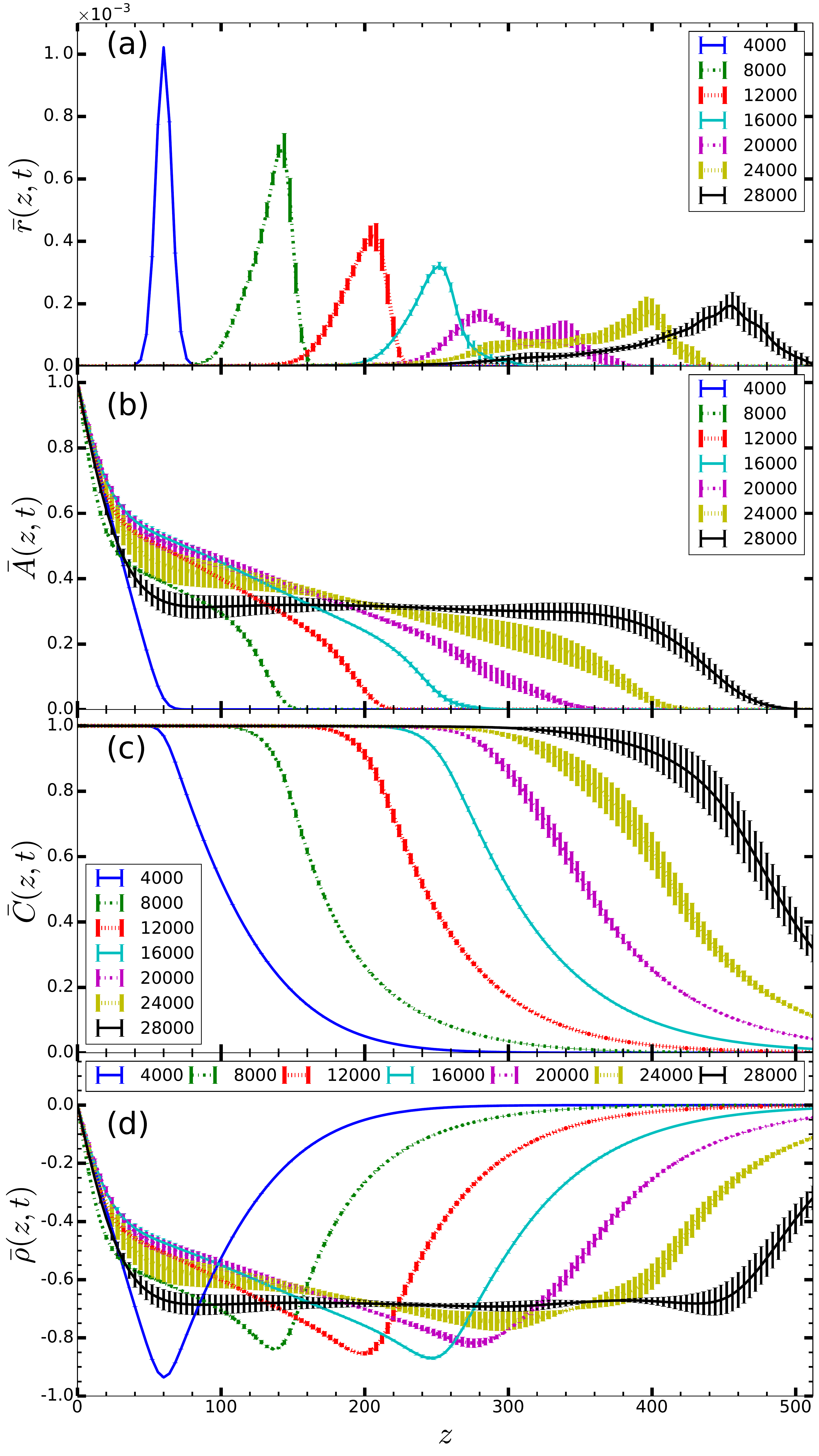}
\caption{Horizontally-averaged profiles of (a) reaction rate $\bar{r}(z,t)$ \labelcref{eq:rbar}, concentration \labelcref{eq:concbar} of (b) dissolving species $\bar{A}(z,t)$ and (c) product $\bar{C}(z,t)$, and (d) density $\bar{\rho}(z,t) = R_A\bar{A}(z,t)+\Delta R_{CB}\bar{C}(z,t)$ at different times $t$ for case R1 illustrated in Figs. \ref{densityField_rc-1}-\ref{concentrationField_rc-1}.
}
\label{zprofiles_rc-1}
\end{figure}

To analyze how the reaction zone evolves in time, we compute the horizontally-averaged reaction rate profile as 
\begin{equation}\label{eq:rbar}
\bar{r}(z,t) = 1/L\int_{0}^{L} A(x,z,t)B(x,z,t) \diff x.
\end{equation}
The reaction profiles shown in Fig. \ref{zprofiles_rc-1}a illustrate that the localized reaction zone enlarges and moves downwards in time.
In parallel, the value of the maximum reaction rate decreases progressively. 
The reaction zone initially consists of a single peak that progressively deforms (see for instance $t=20000$ where two local maxima are visible). 
This can be explained by the intense coalescence occurring around that time: fingers have different lengths as merging fingers are longer than their neighbours (Fig. \ref{densityField_rc-1_200}). 
The reaction zone is initially symmetric, but this symmetry is lost in time as its tail enlarges more than its head, where fingers are arrested by the minimum of density. 

All these characteristics of the reaction zone can be linked to those of the horizontally-averaged concentration profiles, computed as
\begin{equation}\label{eq:concbar}
\bar{f}(z,t) = 1/L\int_{0}^{L} f(x,z,t) \diff x, \quad \text{with $f = A, B, C$,}
\end{equation}
and illustrated at different times in Fig. \ref{zprofiles_rc-1}b-c.
At early times < 8000, the $\bar{A}$ profile (Fig. \ref{zprofiles_rc-1}b) looks like its RD counterpart \cite{loodts15,loodts14prl,loodts16} as convection is not large enough to significantly affect the concentration profiles, while later $\bar{A}$ is deformed due to the apparition of fingering. 
At $t=28000$, we observe a plateau between $z\approx 80$ and 420 in which $\bar{A}\approx 0.3$.
On a given distance, $\bar A$ remains almost constant as fingers contain the same quantity of A, but are thinner near the interface and more spread out just above the minimum of density, which acts as a barrier that prevents fingers from progressing further downwards in the solution (Fig. \ref{aField_rc-1}). 
The plateau of $\bar{A}$ ends where $\bar B = \beta - \bar C $ starts to increase (i.e. $\bar C$ starts to decrease, see Fig. \ref{zprofiles_rc-1}c) as species A is consumed by the reaction and $\bar{r}$ rises (Fig. \ref{zprofiles_rc-1}a).
Lower in the solution, $\bar {r}$ reaches a maximum before decreasing as A has not diffused far enough and has been depleted by the reaction with B (Fig. \ref{zprofiles_rc-1}a). 
Similarly to the A profile, the C profiles initially look like their RD counterparts \cite{loodts14prl,loodts15,loodts16}, i.e. error function-like curves which decrease from their maximum value $\beta=1$ at the reaction front to zero in the bulk of the solution (Fig. \ref{zprofiles_rc-1}c). 
However, no bumps appear in $\bar{C}$ because the fingering pattern is mainly due to the denser dissolving species A (Fig. \ref{concentrationField_rc-1}).
Analyzing the density profile $\bar{\rho}(z,t)$ plotted in Fig. \ref{zprofiles_rc-1}d shows that the width of the minimum of density enlarges progressively, so that it transitions in time from a strict local minimum to a zone where the density is constant, corresponding to the plateau value in $\bar A$ at $t=28000$ (Fig. \ref{zprofiles_rc-1}a). 

\subsubsection{R2 case: more unstable reactive system ($R_A=1$, $\Delta R_{CB}=1$)}
\begin{figure}[htbp]\centering
\subfloat[column 1][$t=800$]{\includegraphics[width=0.49\columnwidth]{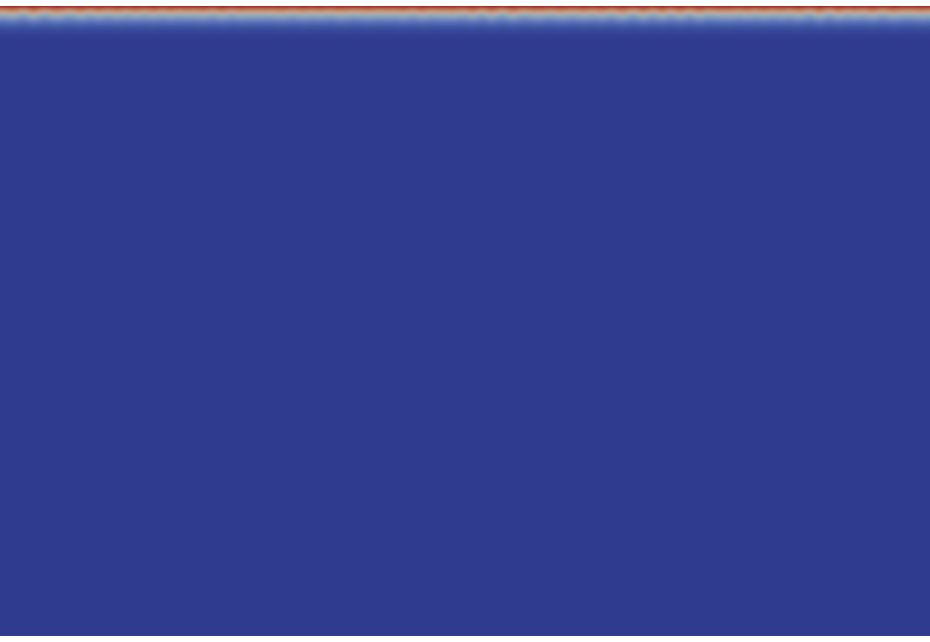}\label{densityField_rc1_6}}\hfill
\subfloat[column 2][$t=1200$]{\includegraphics[width=0.49\columnwidth]{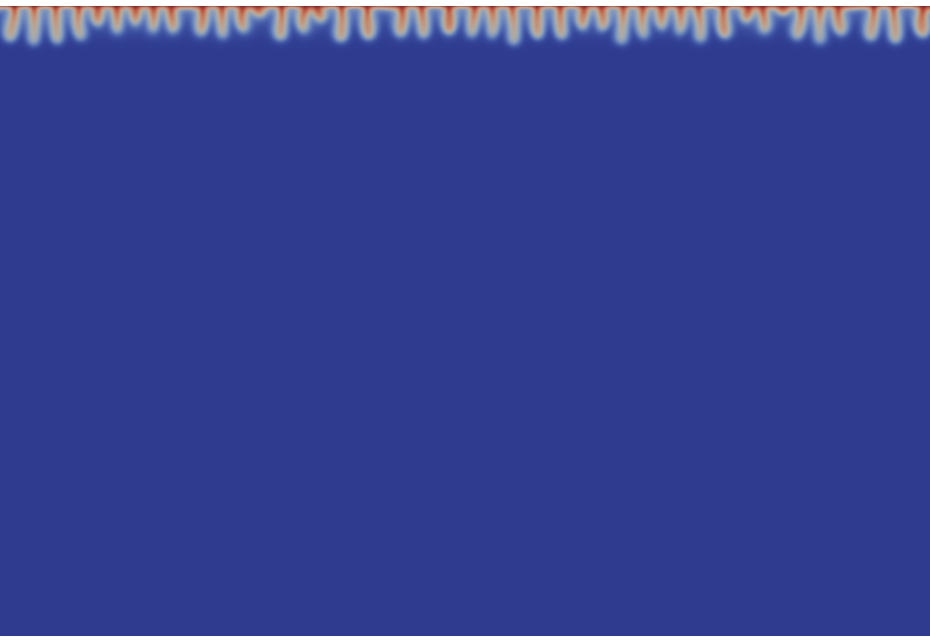} \label{densityField_rc1_10}}\\[-2.5ex]
\subfloat[column 1][$t=2000$]{\includegraphics[width=0.49\columnwidth]{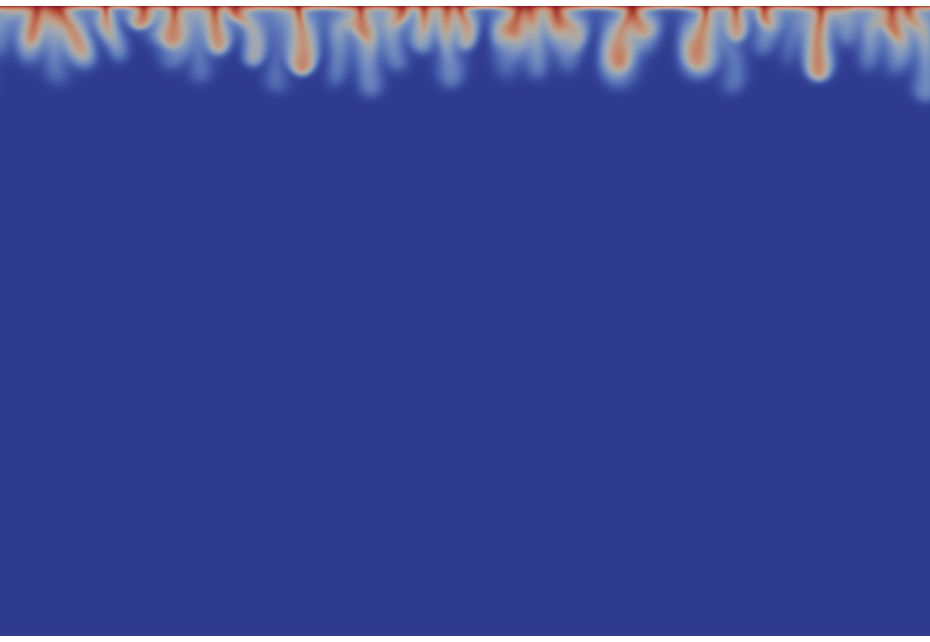} \label{densityField_rc1_20}}\hfill
\subfloat[column 2][$t=4000$]{\includegraphics[width=0.49\columnwidth]{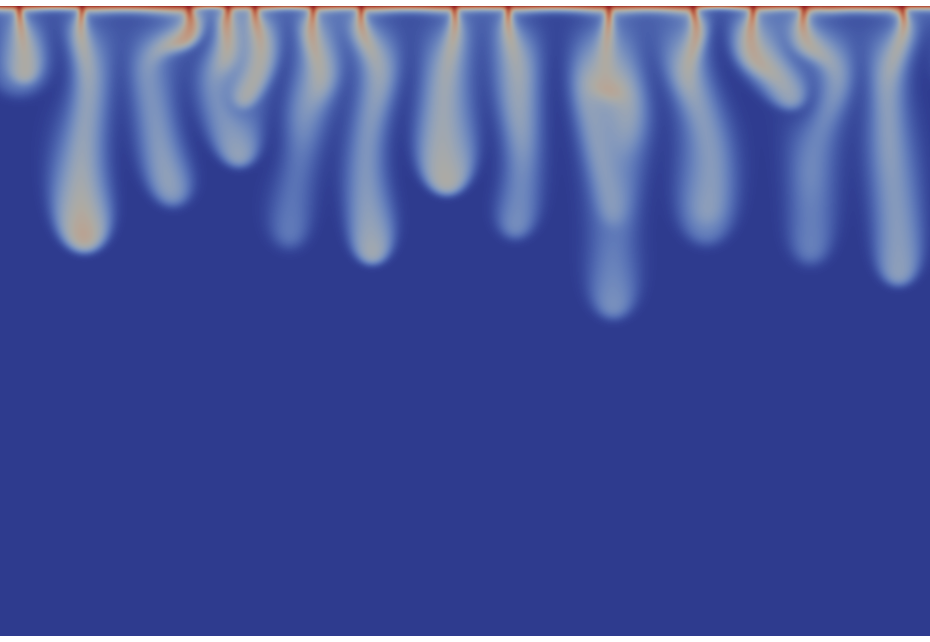}\label{densityField_rc1_50}}\\[-2.5ex]
\subfloat[column 1][$t=8000$]{\includegraphics[width=0.49\columnwidth]{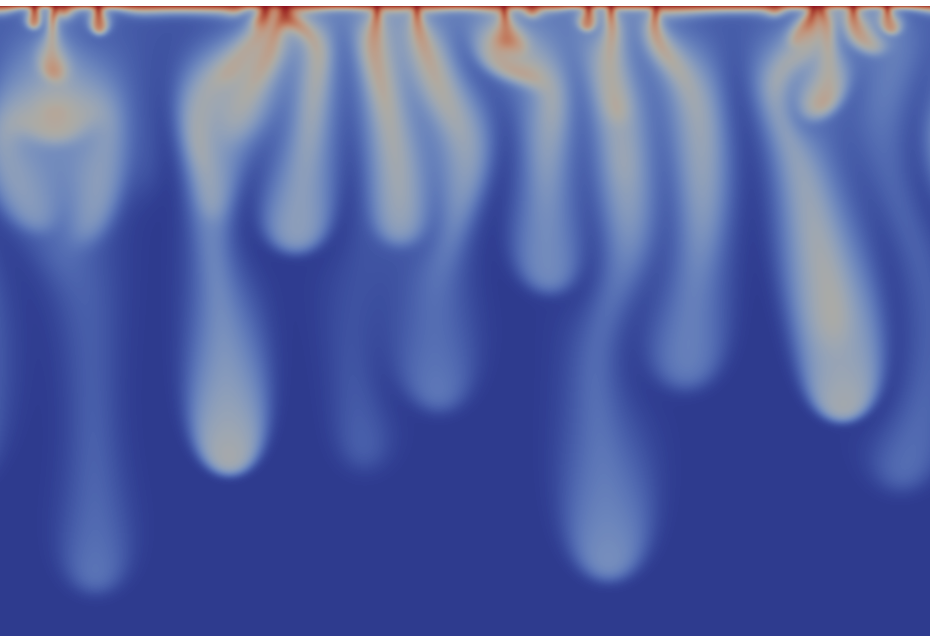}\label{densityField_rc1_100}}\hfill
\subfloat[column 2][$t=12000$]{\includegraphics[width=0.49\columnwidth]{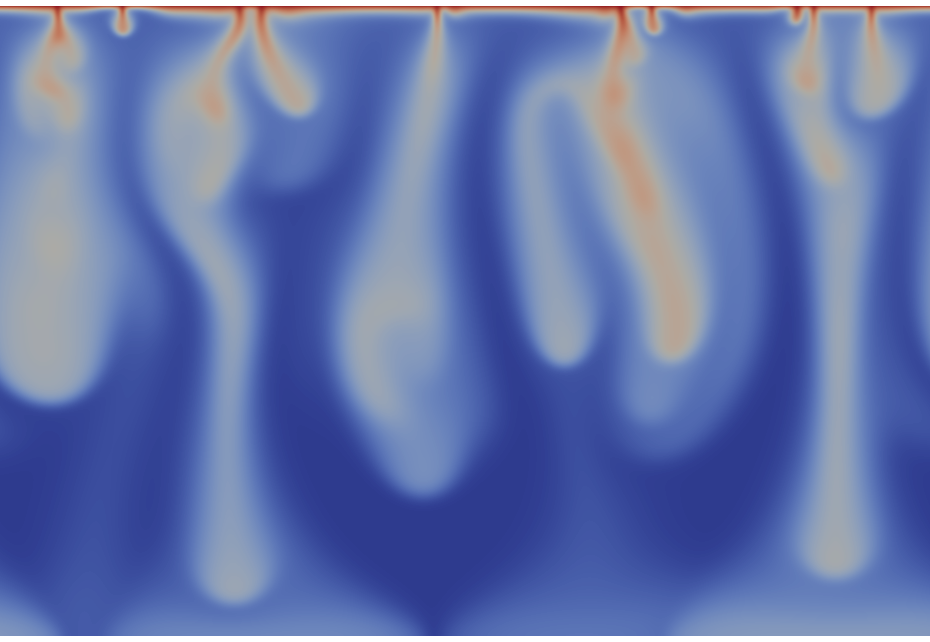}\label{densityField_rc1_150}}
\caption{Density field at different times $t$ in R2 case (destabilizing chemistry, $R_A=1$ and $\Delta R_{CB}=1$). The scale varies between 0 (blue) and 2 (red). }
\label{densityField_rc1}
\end{figure}

When the contribution of the product C to the density is sufficiently larger than that of the dissolved reactant B, reactions destabilize the system even more, i.e. increase the characteristic growth rate in the linear regime \cite{budroni14, loodts14prl, loodts15, budroni17}.
Similarly, fingers develop more quickly and elongate more rapidly than in the NR case (Fig. \ref{densityField_rc1}).
The fingering dynamics in case R2 are characterized by the same successive regimes as in case NR: no fingering (Fig. \ref{densityField_rc1_6}), linear finger growth (Fig. \ref{densityField_rc1_10}), merging (Figs. \ref{densityField_rc1_20}-\ref{densityField_rc1_50}), and reinitiation (Figs. \ref{densityField_rc1_100}-\ref{densityField_rc1_150}).
The number of fingers is initially larger than for NR case (compare Fig. \ref{densityField_rc1_10} with Fig. \ref{density field nr50}), in agreement with previous experimental studies \cite{cherezov16, thomas16, budroni14, budroni17, loodts14prl} and theoretical studies \cite{budroni14, loodts15, budroni17}.

\begin{figure}[htbp]\centering
\subfloat[][$A(x,z,t)$]{\includegraphics[width=0.49\columnwidth]{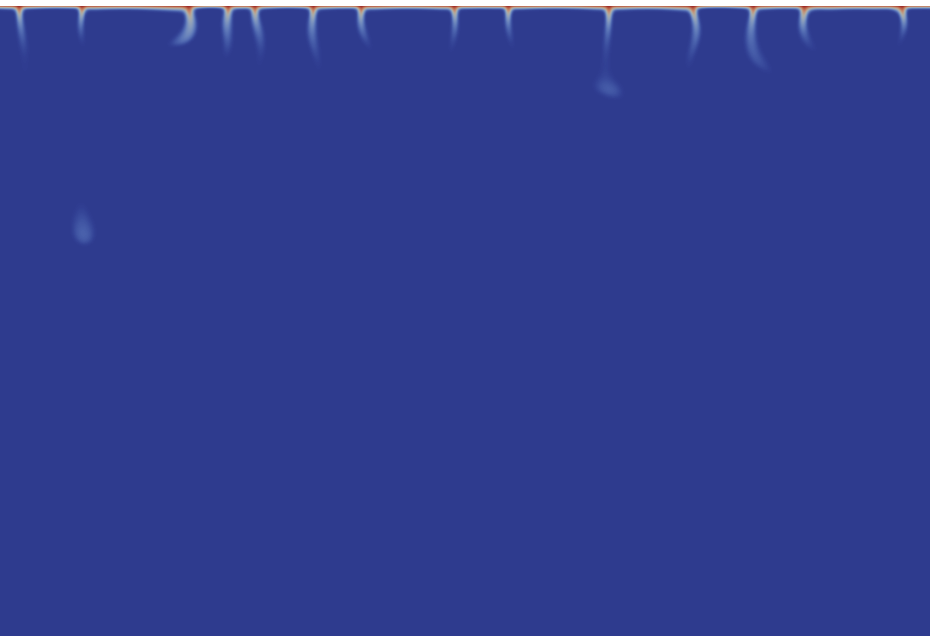}\label{aField_rc1}}\hfill
\subfloat[][$B(x,z,t)$]{\includegraphics[width=0.49\columnwidth]{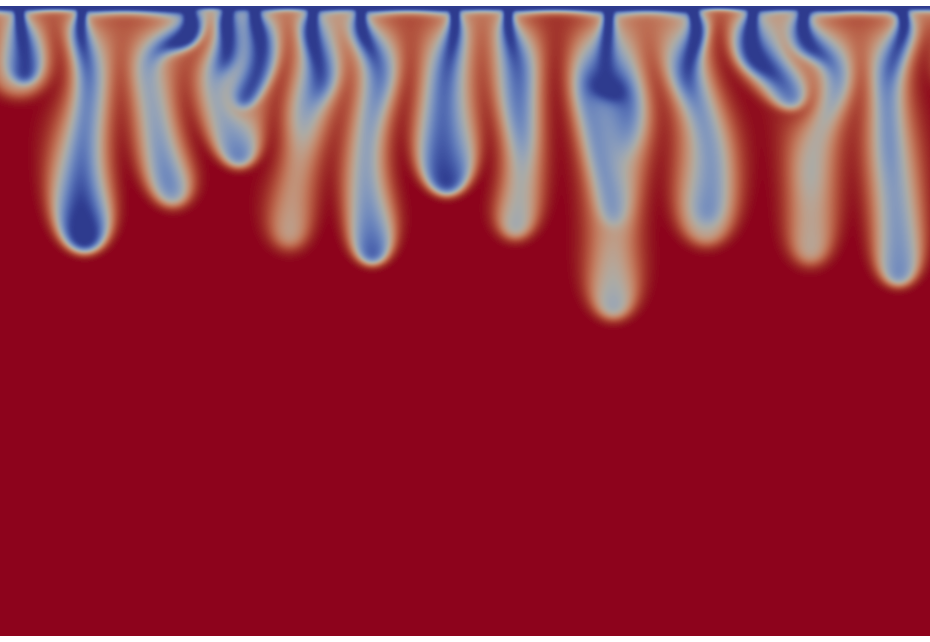}\label{bField_rc1}} \\[-2.5ex]
\subfloat[][$C(x,z,t)$]{\includegraphics[width=0.49\columnwidth]{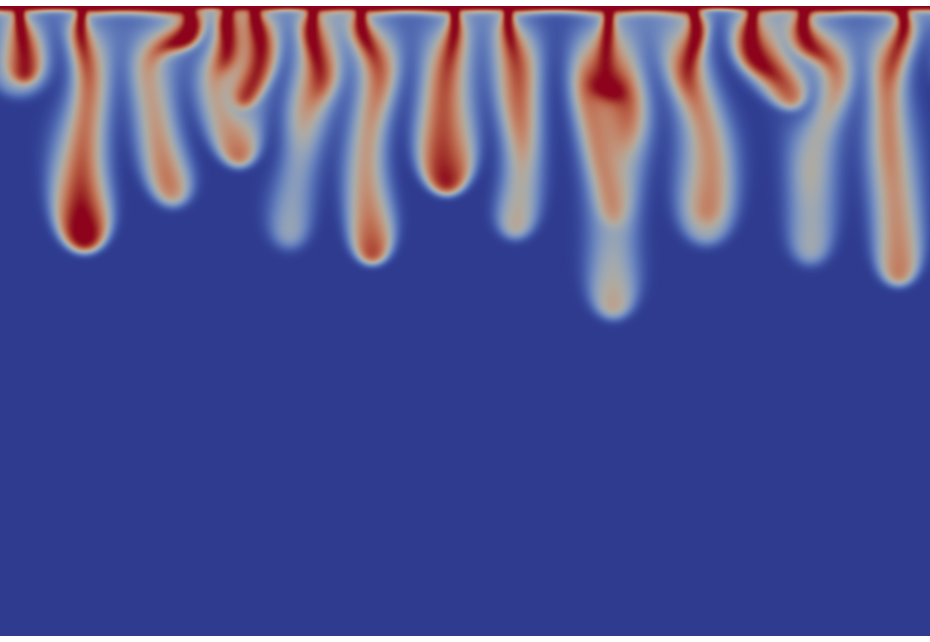}\label{cField_rc1}}\hfill
\subfloat[][$AB(x,z,t)$]{\includegraphics[width=0.49\columnwidth]{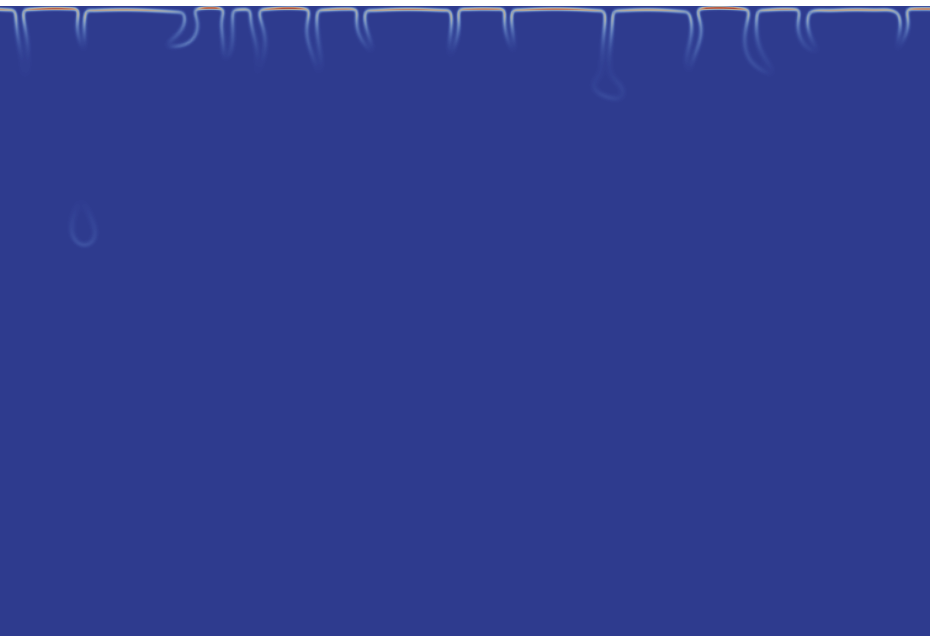} \label{abField_rc1}}
\caption{Concentration fields of the dissolving species A (a), reactant B (b) and product C (c) varying between 0 (blue) and 1 (red), and reaction rate $AB$ (d) varying between 0 (blue) and 0.01 (red) at time $t=4000$ corresponding to the density field shown in Fig. \ref{densityField_rc1_50} for case R2.}
\label{concentrationField_rc1}
\end{figure}

We now analyze in Fig. \ref{concentrationField_rc1} the concentration fields at a given time to understand how they combine to form the fingering pattern observed in the density field (Fig. \ref{densityField_rc1_50}).
The dissolving species A does not penetrate far in the host solution (Fig. \ref{aField_rc1}). 
The reactant A is indeed readily consumed as soon as it enters the host phase, which limits its progression into the host solution, in agreement with recent results \cite{cherezov16, ghoshal17}.
The reaction must be fed by B diffusing to the upper part of the solution, close to the interface, where it reacts with dissolved A (Fig. \ref{bField_rc1}).
The fingering pattern observed in the density field (Fig. \ref{densityField_rc1_50}) is mainly due to the denser product C produced close to the interface, sinking into the lower part of the host solution (Fig. \ref{cField_rc1}).
The convection mechanism is thus essentially the same as in the non-reactive counterpart (NR case), except that the contribution of the denser product C adds to that of the dissolving species A.
This explains why the same types of regimes are observed in both cases.
The reaction rate $AB$ is the largest along the side of the birth zone of the density fingers and the contact line between the boundary layer rich in A and the bulk solution (Fig. \ref{abField_rc1}). 

\subsubsection{R3 case: unstable due to reaction ($R_A=-1$, $\Delta R_{CB}=1$)}
\begin{figure}[htbp]\centering
\subfloat[][$t=2000$]{\includegraphics[width=0.49\columnwidth]{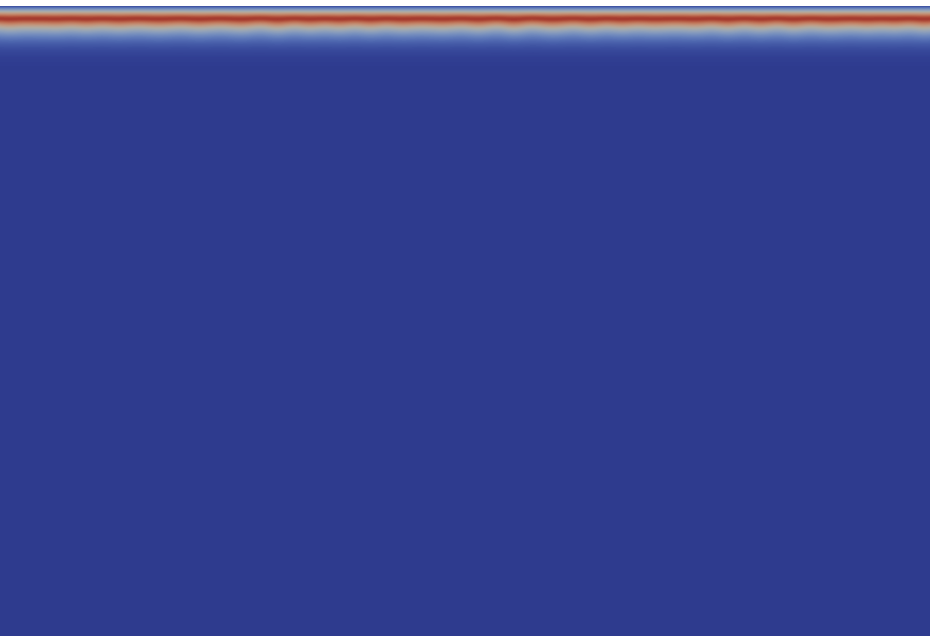}\label{densityField_ra-1_rc1_10}}\hfill
\subfloat[][$t=4000$]{\includegraphics[width=0.49\columnwidth]{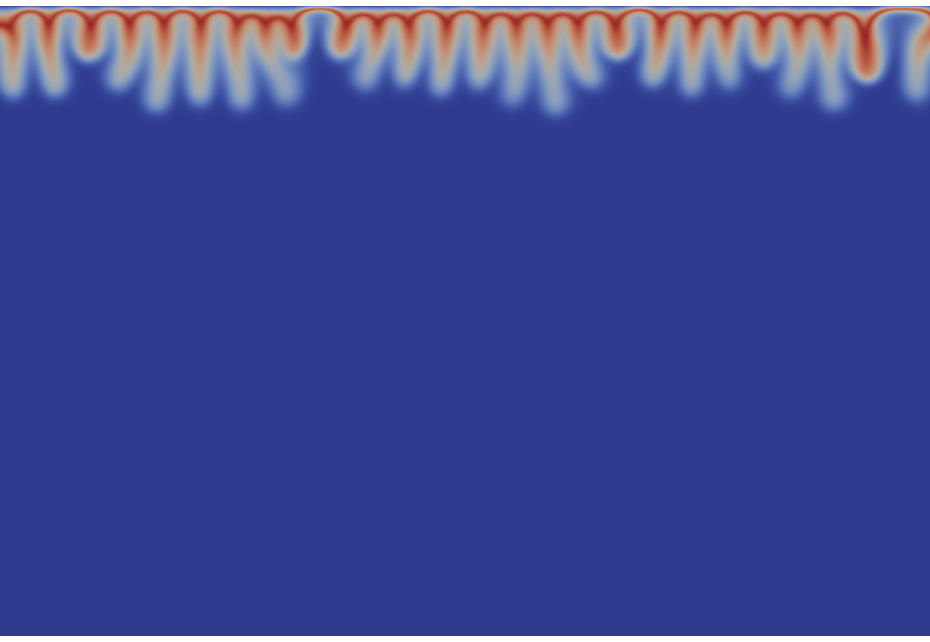} \label{densityField_ra-1_rc1_30}}\\[-2.5ex]
\subfloat[][$t=6000$]{\includegraphics[width=0.49\columnwidth]{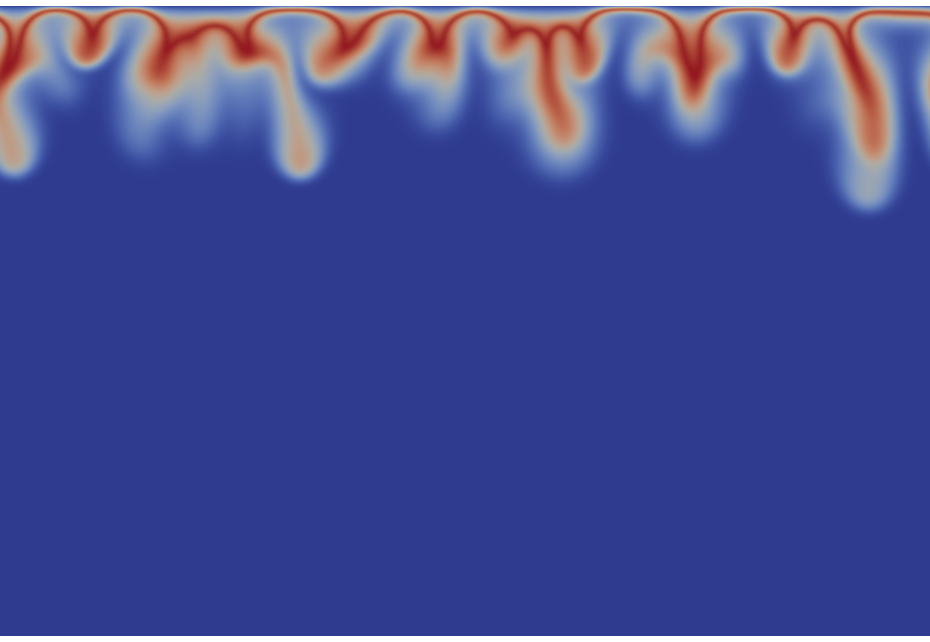}\label{densityField_ra-1_rc1_50}}\hfill
\subfloat[][$t=8000$]{\includegraphics[width=0.49\columnwidth]{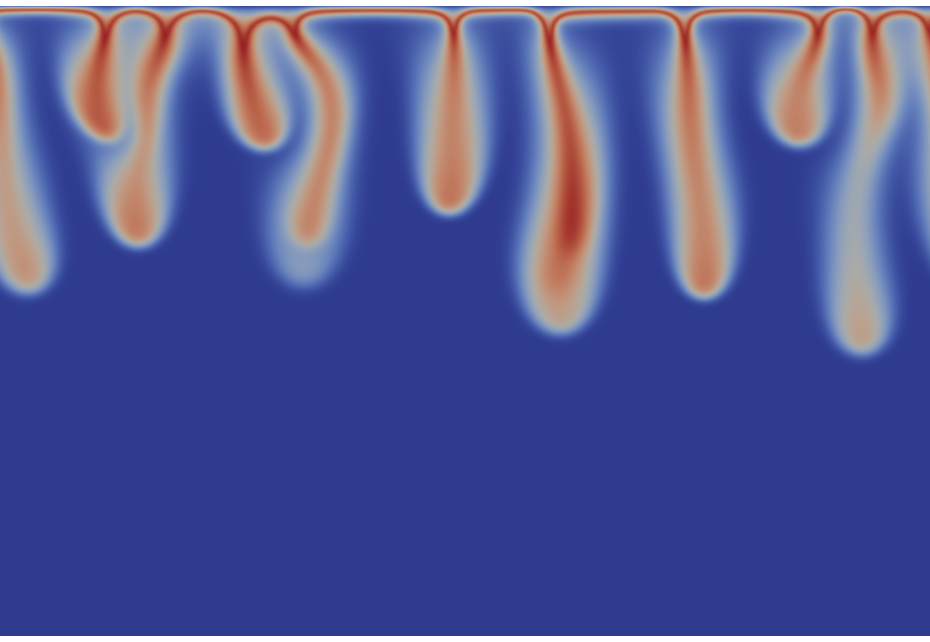}\label{densityField_ra-1_rc1_100}}\\[-2.5ex]
\subfloat[][$t=12000$]{\includegraphics[width=0.49\columnwidth]{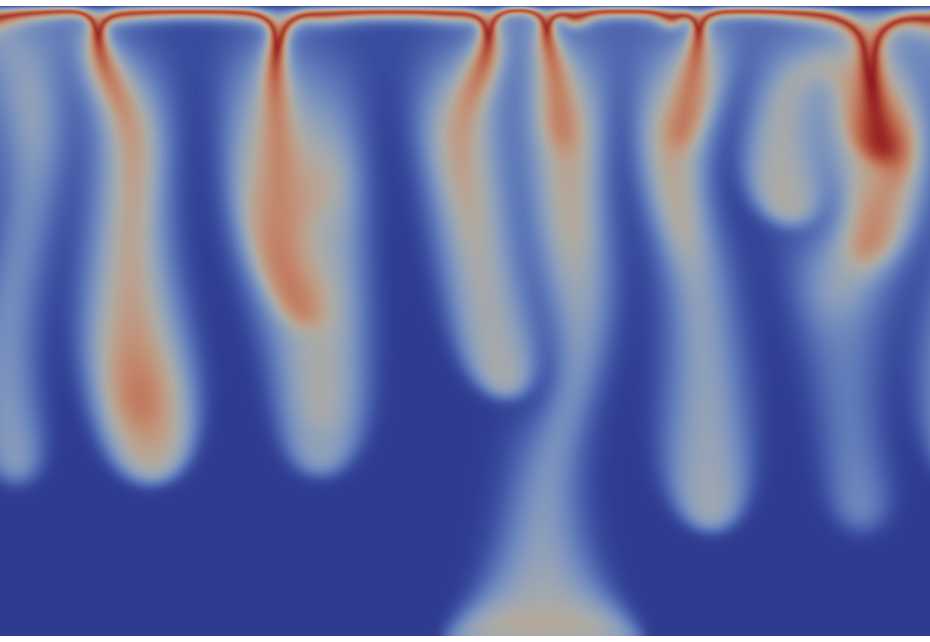}\label{densityField_ra-1_rc1_150}}\hfill
\subfloat[][$t=20000$]{\includegraphics[width=0.49\columnwidth]{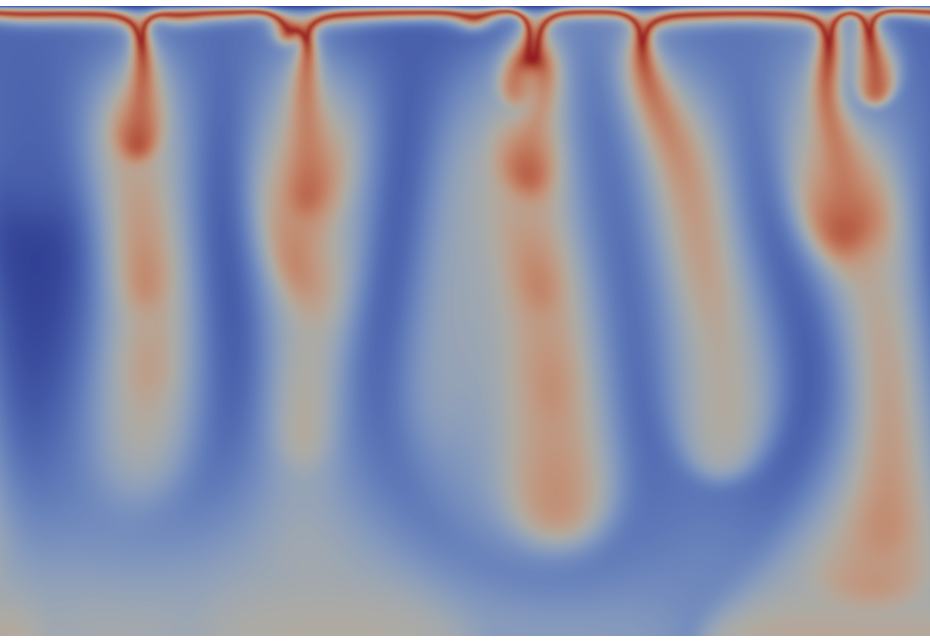}\label{densityField_ra-1_rc1_200}}
\caption{Density field at different times $t$ in R3 case (unstable due to reaction, $R_A=-1$ and $\Delta R_{CB}=1$). 
	The scale varies between 0 (blue) and 1 (red). 
	}
\label{densityField_ra-1_rc1}
\end{figure}

When $R_A<0$, the non-reactive case is stable and A+B $\rightarrow$ C reactions can be at the origin of buoyancy-driven convection as soon as $\Delta R_{CB} > 0$ because a maximum, corresponding to a locally unstable stratification, then develops in the density profile \cite{loodts15,loodts16}. 
Therefore we expect a stable boundary layer just below the interface followed further away at a given distance from the interface by a locally unstable zone generating fingers sinking down.
Our numerical results confirm this prediction as shown in Fig. \ref{densityField_ra-1_rc1}. 
Fingers develop from the maximum of density located below the interface and their base has a characteristic shape not observed in the other cases.
Apart from that, the dynamics are similar to those observed in case R2 (Fig. \ref{densityField_rc1}): no fingering (Fig. \ref{densityField_ra-1_rc1_10}), linear finger growth (Fig. \ref{densityField_ra-1_rc1_30}), merging (Figs. \ref{densityField_ra-1_rc1_50}-\ref{densityField_ra-1_rc1_100}), and reinitiation  (Figs. \ref{densityField_ra-1_rc1_150}-\ref{densityField_ra-1_rc1_200}).
The reinitiation mechanism can be observed in Fig. \ref{densityField_ra-1_rc1_200} as ``pulses'' in the fingers, corresponding to merging with protoplumes.

\begin{figure}[htbp]\centering
\subfloat[][$A(x,z,t)$]{\includegraphics[width=0.49\columnwidth]{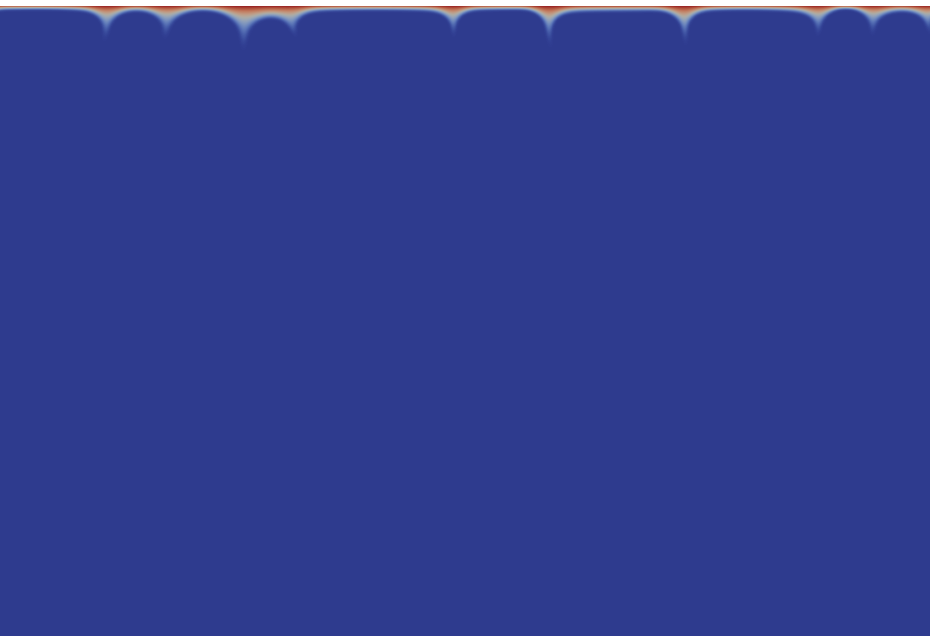}\label{aField_ra-1_rc1}}\hfill
\subfloat[][$B(x,z,t)$]{\includegraphics[width=0.49\columnwidth]{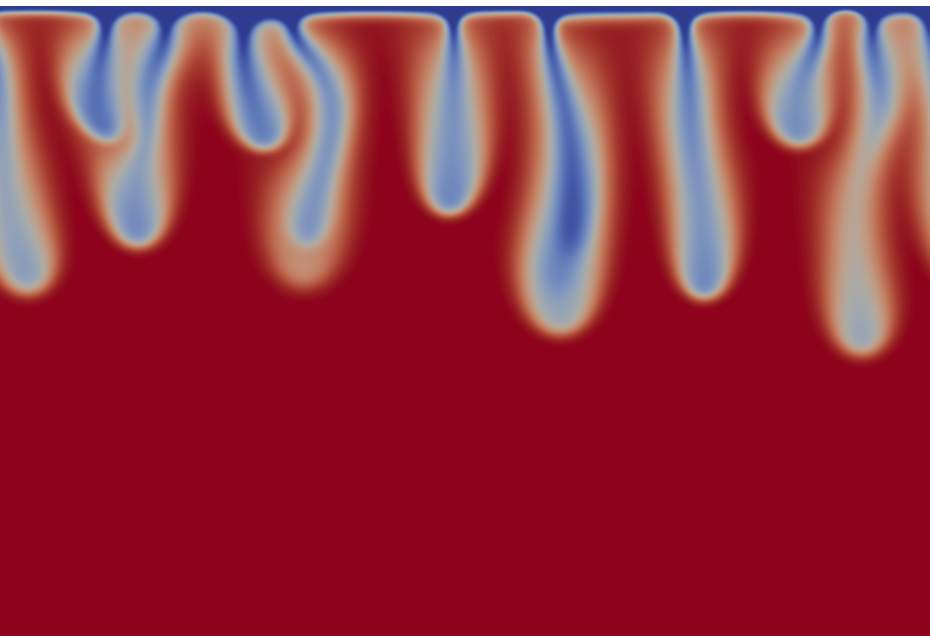}\label{bField_ra-1_rc1}}\\[-2.5ex]
\subfloat[][$C(x,z,t)$]{\includegraphics[width=0.49\columnwidth]{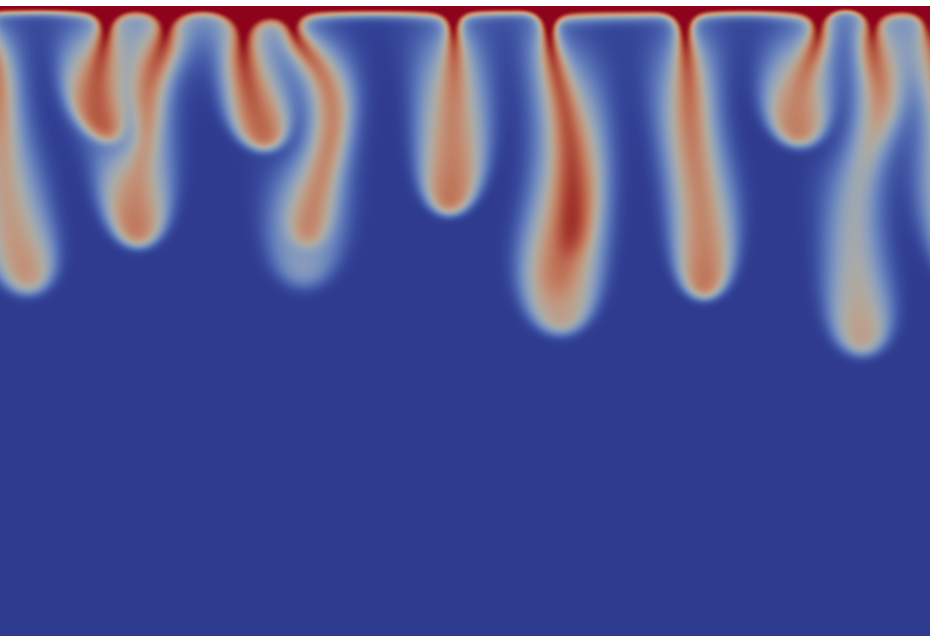}\label{cField_ra-1_rc1}} \hfill
\subfloat[][$AB(x,z,t)$]{\includegraphics[width=0.49\columnwidth]{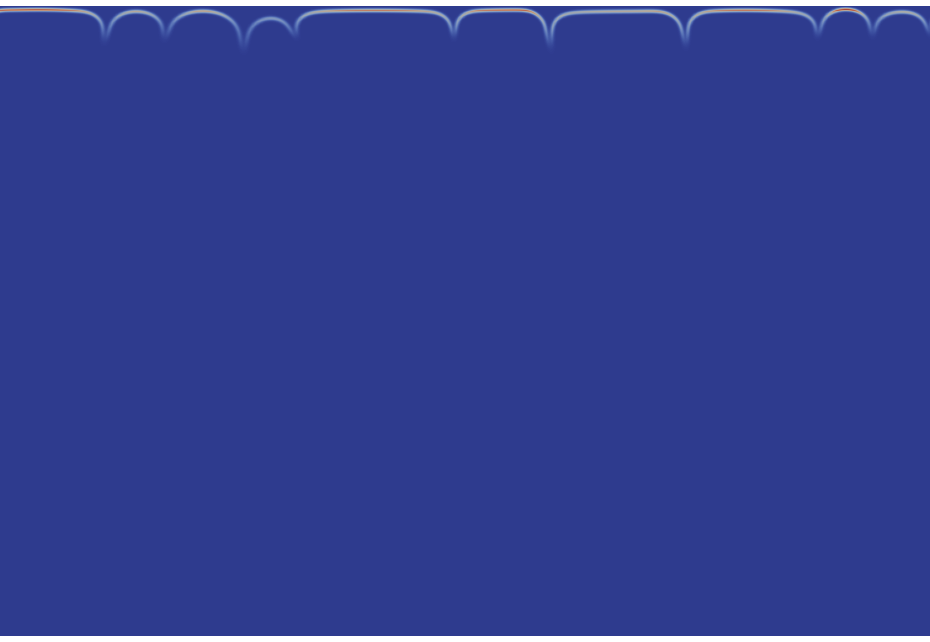} \label{abField_ra-1_rc1}} 
\caption{Concentration fields of the dissolving species A (a), reactant B (b) and product C (c) varying between 0 (blue) and 1 (red), and reaction rate $AB$ (d) varying between 0 (blue) and 0.008 (red) at time $t=8000$ corresponding to the density field shown in Fig. \ref{densityField_ra-1_rc1_100} for case R3.}
	\label{concentrationField_ra-1_rc1}
\end{figure}

Like in R2 case, the fingering pattern (Fig. \ref{densityField_ra-1_rc1_100}) is also mostly formed by solute C (Fig. \ref{cField_ra-1_rc1}) sinking into the less dense solution of B (Fig. \ref{bField_ra-1_rc1}) while most of the dissolving species A is consumed as soon as it enters the solution (Figs. \ref{aField_ra-1_rc1}-\ref{abField_ra-1_rc1}).
The remaining A on top decreases the density of the solution and is thus at the origin of the specific shape of the fingers' base. 
Except that specific fingers' shape, the dynamics is thus qualitatively similar but slower than that in R2 case, probably due to the stable boundary layer between the interface and the maximum of density. 

\subsection{Comparison of the space-time plots}\label{space-time sec}
\begin{figure}[htbp]\centering
\includegraphics[width=\columnwidth]{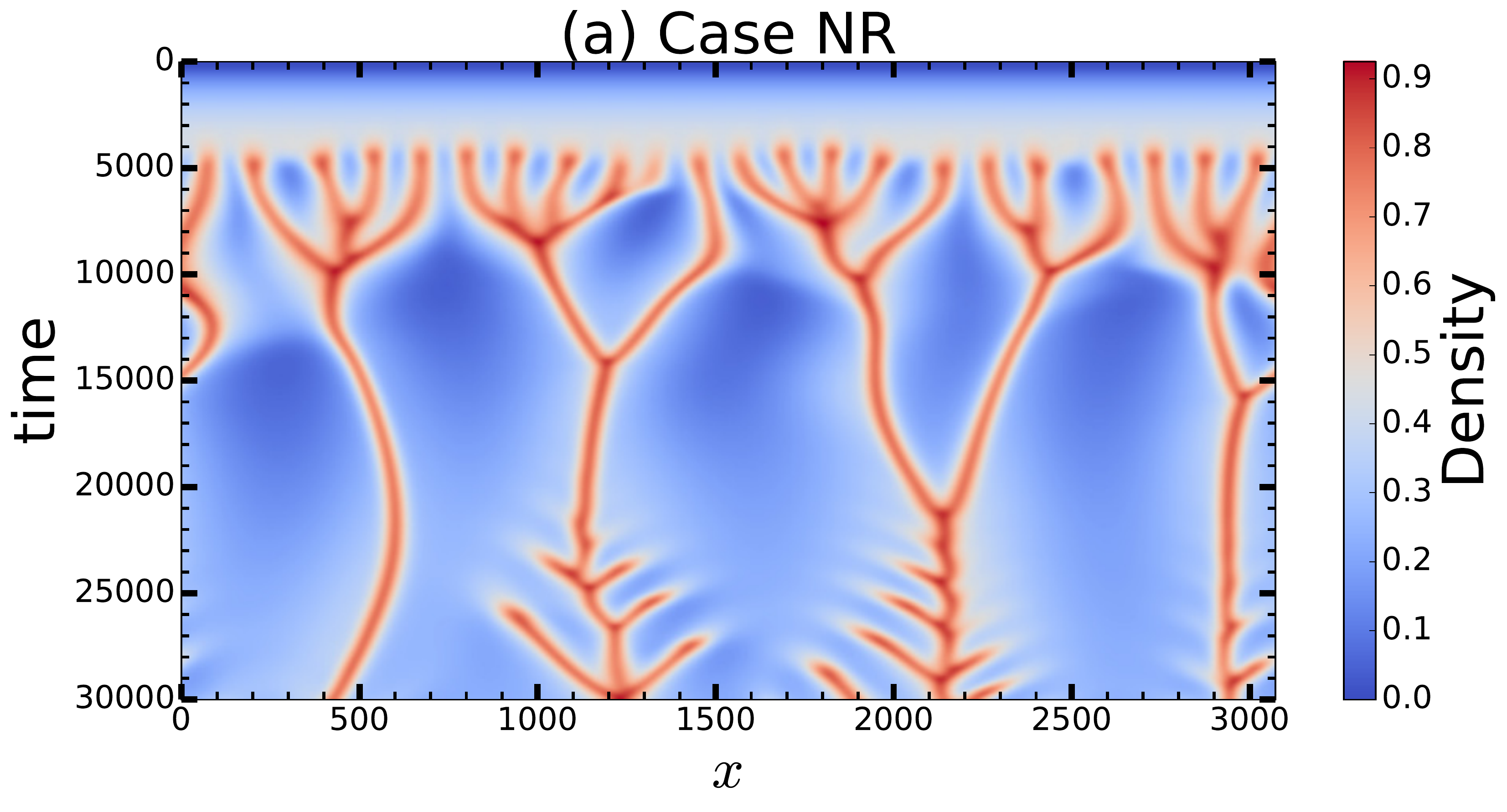}\\
\includegraphics[width=\columnwidth]{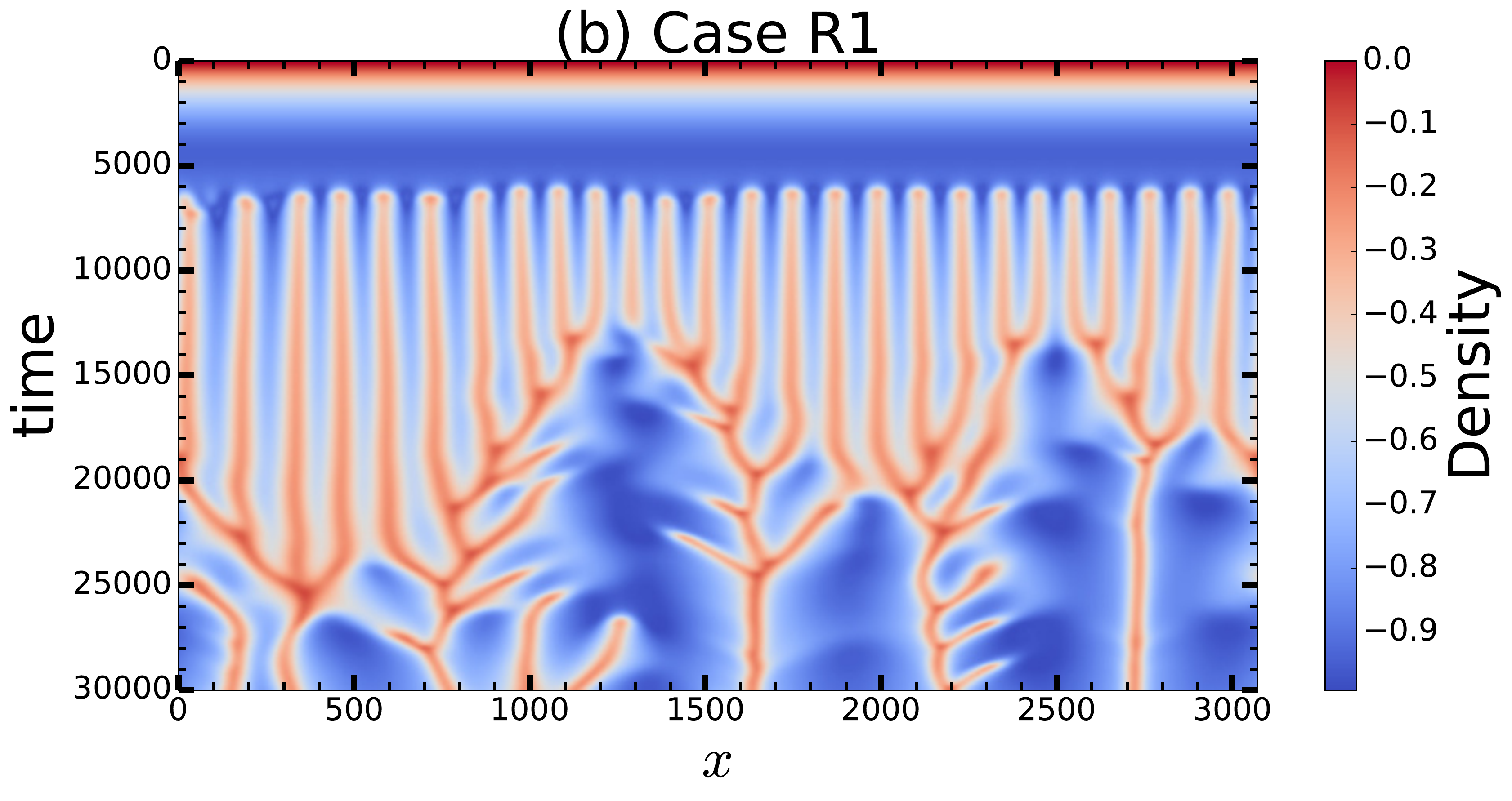}\\
\includegraphics[width=\columnwidth]{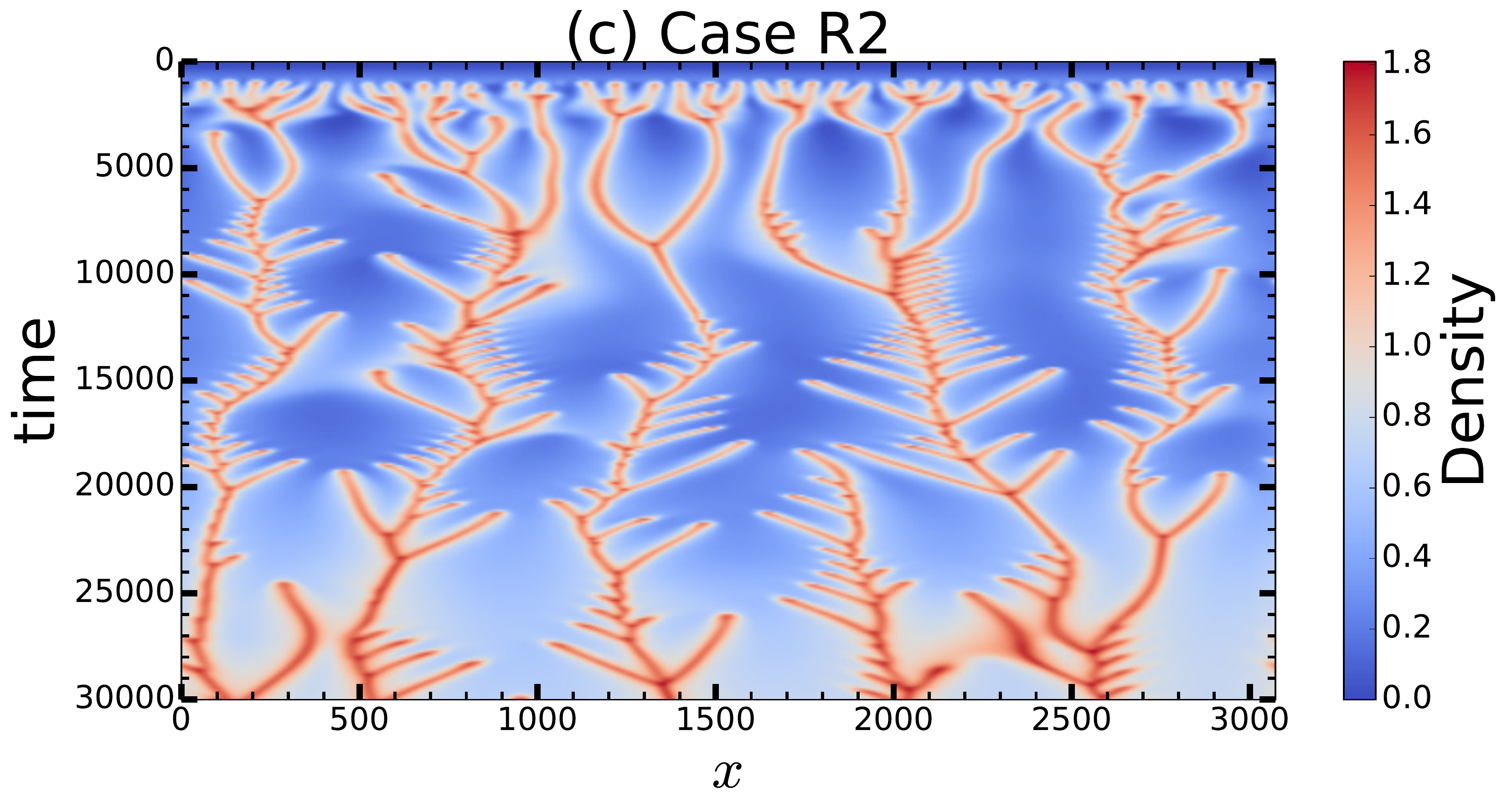}\\
\includegraphics[width=\columnwidth]{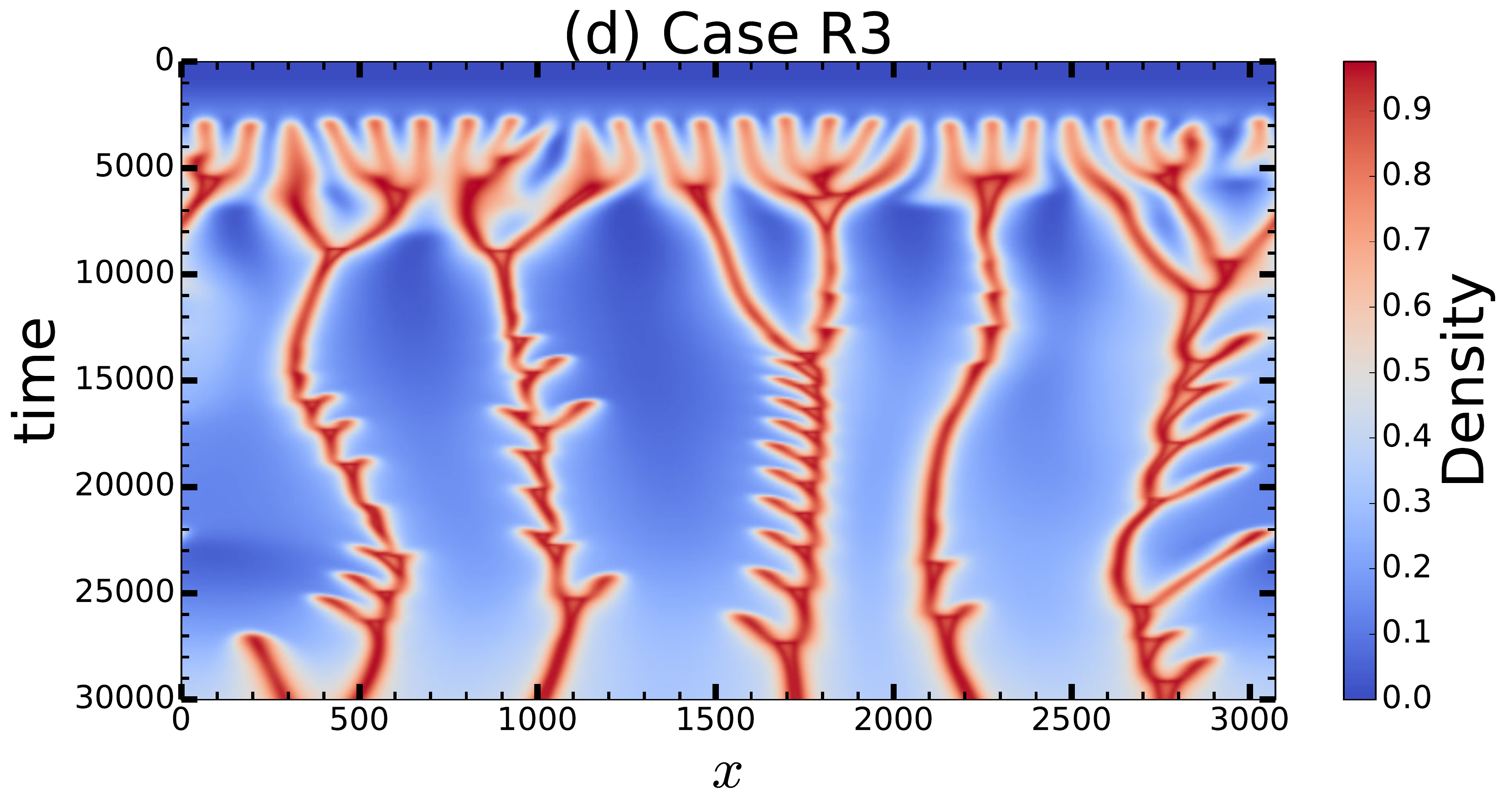} 
\caption{Space-time plots showing the dynamics of the finger roots, constructed by plotting the density along the horizontal line at $z=64$ (except for (d) at $z=128$) as a function of time, for (a) NR case illustrated in Fig. \ref{density field nr}, (b) R1 case in Figs. \ref{densityField_rc-1}-\ref{zprofiles_rc-1}, (c) R2 case in Figs. \ref{densityField_rc1}-\ref{concentrationField_rc1} and (d) R3 case in Figs. \ref{densityField_ra-1_rc1}-\ref{concentrationField_ra-1_rc1}.}
\label{space-time}
\end{figure}

The space-time plots shown in Fig. \ref{space-time} summarize the differences between the dynamics in the NR case  and those in the three specific reactive cases R1, R2 and R3 discussed above. 
These pictures are constructed by plotting the density along a horizontal line at $z=64$, except for case R3 where we had to take a line below at $z=128$ because of the stable boundary layer close to the interface. 
The dynamics in case R1 (Fig. \ref{space-time}b) are slower while those in case R2 (Fig. \ref{space-time}c) are faster than in case NR (Fig. \ref{space-time}a) because of the different density profiles in the host phase.
In particular, fingers appear earlier and new protoplumes are generated more frequently in case R2 than in case NR, and vice versa for case R1. 
In case R3, fingering occurs (Fig. \ref{space-time}d) because of the effect of the reaction on the density profile.
The successive regimes describing the dynamics are similar for cases NR, R2 and R3, while in case R1, some reinitiation already starts during the merging regime. 
For all reactive cases R1, R2 and R3, the wavelength of the fingering pattern when fingers first become visible is smaller than for case NR, in agreement with theoretical and experimental predictions \cite{loodts14prl,loodts15,budroni14,budroni17, cherezov16,thomas16}.

\section{Effect of reaction-diffusion-convection interplay on the spatio-temporal dynamics}\label{dynamics}
Motivated by the differences between the dynamics of cases NR, R1, R2 and R3, we further analyze the spatio-temporal dynamics of fingering and of reaction when changing $\Delta R_{CB}$ between -1 and 1. 
Our aim is to quantify how the reaction affects the convective dynamics and conversely how convection impacts the dynamics of the reaction zone. 
We have previously classified the effects of reaction on convection by evaluating a characteristic growth rate in the linear regime \cite{loodts14prl,loodts15}. 
Below a given value $\Delta_R$ for $\Delta R_{CB}$, this growth rate is smaller than its non-reactive counterpart, and conversely. 
We now revisit this classification by examining various aspects of the convective dynamics in the fully developed non-linear regime before shutdown: firstly in Fig. \ref{dynamics} the spatio-temporal dynamics, i.e. the evolution of the fingering pattern (elongation, wavelength) and of the reaction zone, and secondly in Fig. \ref{storage rate} the storage properties, i.e. at what rate and under which form A is stored into the host fluid phase.
Note that the trends observed when varying $\Delta R_{CB}$ for a given $R_A$, i.e. when changing the reactant B -- product C pair for a given dissolving species A, are similar for $R_A>0$ and $R_A<0$.
These trends will therefore be illustrated for $R_A=1$ only as the same discussion can be repeated for $R_A=-1$.

\subsection{Mixing length and velocity}
We first analyze the impact of changing $\Delta R_{CB}$ on the evolution of the mixing length and on the finger velocity. 
We define the mixing length $z_m$ as the most advanced position along $z$ where $A+C > s$, with $s$ a small arbitrary threshold here chosen as $0.01$. 
This position evolves dynamically in time as A dissolves into the host solution and its reaction with solute B produces C.
This definition of the mixing length represents the extension of the zone containing stored A in the form of either dissolved unreacted A or product C (i.e. reacted A).

We can derive analytical expressions for the mixing length valid before convection sets in.
In the non-reactive case, inserting $A(z_m)=s$ into the diffusive concentration profile \cite{Diffusion,loodts14chaos} ${\rm erfc}(z/(2\sqrt{t}))$ of A gives
	\begin{equation}\label{eq:zm_nr}
	z_m(t) = 2 {\rm erfinv}(1-s) \sqrt{t},
	\end{equation}
$\approx 3.64 \sqrt{t}$ for $s=0.01$.
In the reactive case, we introduce $C(z_m)=s$ into the RD concentration profile \cite{loodts15} $2{\rm erfc}(z/(2\sqrt{t}))$ of C valid for $\beta=1$ below the reaction front to get
	\begin{equation}\label{eq:zm}
	z_m(t) = 2 {\rm erfinv}(1-s/2) \sqrt{t}, 
	\end{equation}
$\approx 3.97 \sqrt{t}$ for $s=0.01$.
Equations \eqref{eq:zm_nr}-\eqref{eq:zm} show that the mixing length in the diffusive regime increases in time as $\sqrt{t}$, and increases faster in the presence of a reaction.

\begin{figure}[htbp]\centering
\includegraphics[width=\columnwidth]{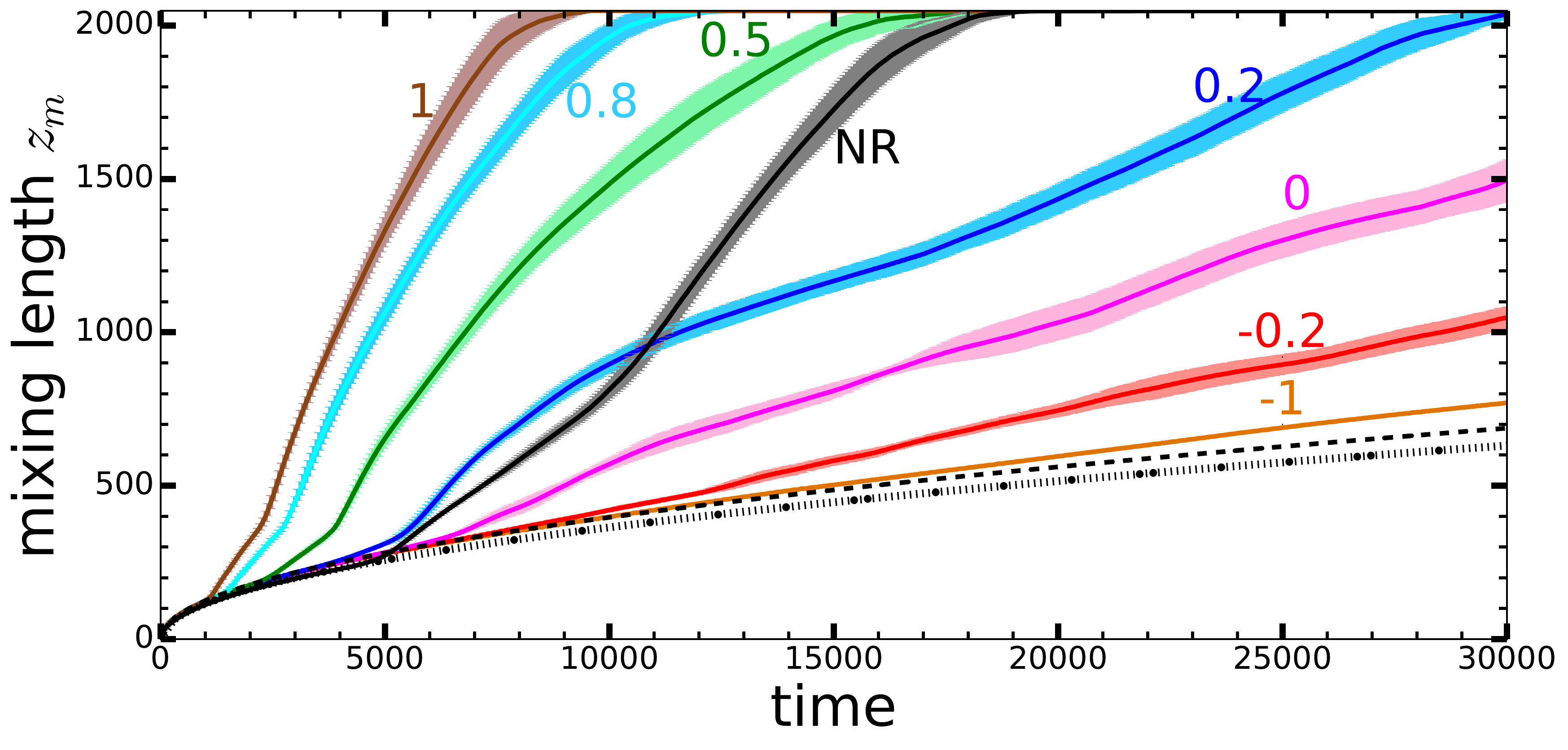}
\caption{Mixing length $z_m$, defined as the most advanced position where $A+C > s=0.01$, as a function of time for  $R_A=1$ and different $\Delta R_{CB}$ indicated in the graph. 
	The dotted and dashed black curves represent the mixing lengths in the diffusive regimes for the non-reactive (Eq. \eqref{eq:zm_nr}) and reactive (Eq. \eqref{eq:zm}) cases, respectively. 
	}
\label{finger length}
\end{figure}

The mixing lengths computed from the numerical simulations are shown in Fig. \ref{finger length}.
One curve represents the average over 15 realizations.
The 95\% confidence interval shown as lighter areas around the curves represents the variability due to the random noise on the initial condition.
We see that the variability between realizations is amplified when $\Delta R_{CB}$ increases, because of more intense convection.

\begin{figure}[tbhp]\centering
\includegraphics[width=\columnwidth]{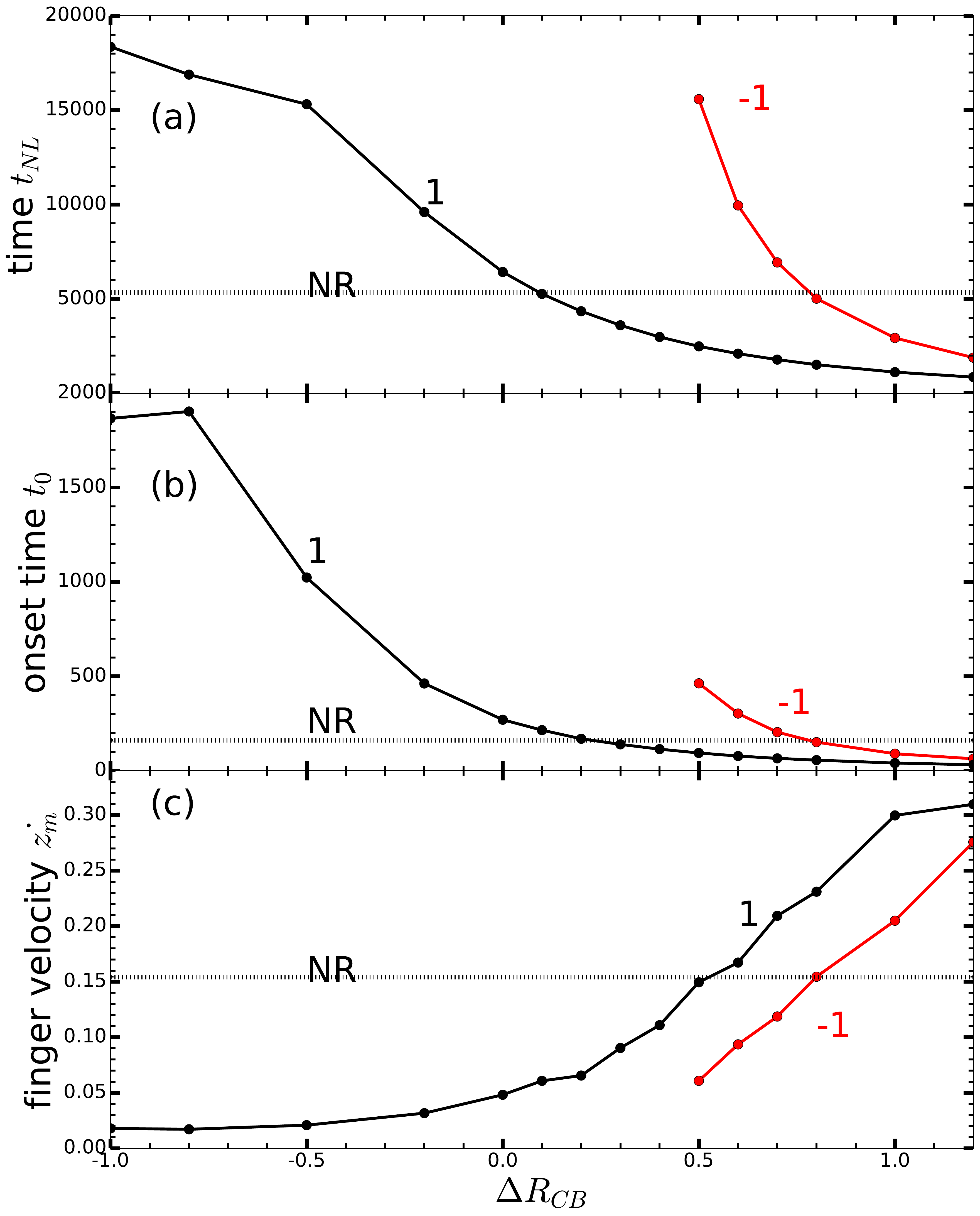}
\caption{(a) Time $t_{NL}$, (b) onset time $t_0$ and (c) finger velocity $\dot{z}_m$ as a function of $\Delta R_{CB}$ for $R_A$ =1 or -1 indicated in the graph. 
The dotted line represents the data for NR case.}
\label{finger velocity}
\end{figure}

Both expressions \eqref{eq:zm} and \eqref{eq:zm_nr} (shown in Fig. \ref{finger length} as dotted and dashed curves respectively) are valid as long as diffusion remains the dominant transport process.
After a certain transition time noted $t_{NL}$, convection becomes important, fingers start to move faster and $z_m$ starts departing from the $\sqrt{t}$ curve.
$z_m$ then increases approximately proportional to $t$, although fingers might progressively slow down in some cases (see e.g. $\Delta R_{CB}=0.5$ after $t=7500$ in Fig. \ref{finger length}), or exhibit two successive different velocities.
$t_{NL}$ represents the time when non-linearities significantly affect the vertical elongation of the fingers. 
More precisely, we evaluate $t_{NL}$ as the time when the relative difference between $z_m$ and the diffusive prediction (Eq. \eqref{eq:zm_nr} for non-reactive cases or Eq. \eqref{eq:zm} for reactive cases) becomes larger than 5\%. 
This time $t_{NL}$ decreases when $\Delta R_{CB}$ increases (Figs. \ref{finger length}, \ref{finger velocity}a), which is coherent with the predictions of the linear stability analysis that the destabilizing effect of chemistry increases when $\Delta R_{CB}$ increases \cite{loodts15}.
For $\Delta R_{CB} \leq 0.1$, $z_m$ deviates from the diffusive curve later than in the non-reactive case.
This value is of the same order as the critical value 0.32 predicted by linear stability analyses \cite{loodts14prl,loodts15}.
The difference might arise because $t_{NL}$ is measured here on the basis of fingering dynamics rather than on the basis of the perturbation with regard to the base state.

\begin{figure}[tbhp]\centering
\includegraphics[width=\columnwidth]{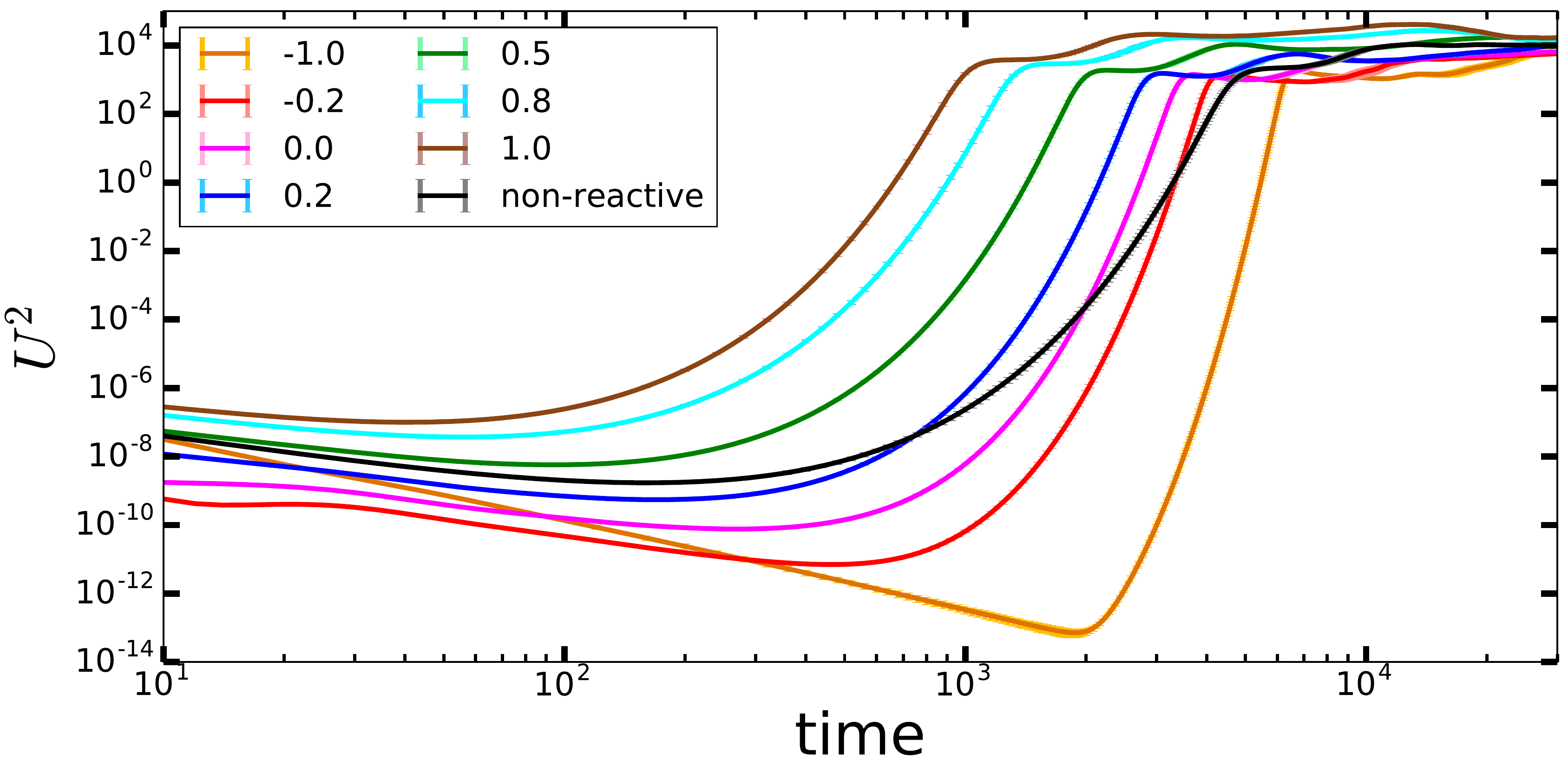}
\caption{Perturbation $U^2$ in velocity for $R_A=1$ and different $\Delta R_{CB}$ indicated in the graph.}
\label{u2}
\end{figure}

We therefore also define a time $t_0$ for the onset of the instability on the basis of the perturbation in velocity, computed as $U^2(t) = \int_{0}^{H}\int_{0}^{L} u_x^2(x,z,t) \diff x \diff z + \int_{0}^{H}\int_{0}^{L} u_z^2(x,z,t) \diff x \diff z$. 
To highlight the dynamics at early times, we plot $U^2(t)$ in a log-log graph (Fig. \ref{u2}).
For any value of $\Delta R_{CB}$, the perturbation initially decreases until a given onset time $t_0$ when it reaches its minimum. 
Like $t_{NL}$, this onset time $t_0$ increases when $\Delta R_{CB}$ is increased (Fig. \ref{finger velocity}b). 
In other words, both linear growth and non-linear regimes start earlier if $\Delta R_{CB}$ is larger. 
The critical value $\Delta R_{CB}=0.2$ above which the reaction makes the system more unstable is slightly larger than the one calculated with $t_{NL}$, but still smaller than the prediction of the linear stability analysis \cite{loodts14prl,loodts15}. 

In addition, we classify the effects of reaction on convection at later times, based on the time-averaged finger velocity $\dot{z}_m$ computed as the least-squares fitted slope of $z_m(t)$ between $t_{NL}$ and the time when the fingers arrive at 95\% of the depth of the host phase. 
$\dot{z}_m$ increases when $\Delta R_{CB}$ increases (Fig. \ref{finger velocity}c), so that as above we can define a critical value of $\Delta R_{CB}$ above which fingers advance faster than their non-reactive counterparts.
This critical value of 0.5 is larger than that evaluated on the basis of $t_{NL}$ and $t_0$.
Indeed for $\Delta R_{CB}=0.5$, although fingers start to accelerate earlier than their non-reactive counterparts (smaller $t_{NL}$), they move more slowly at later times and arrive at the bottom at the same time as in the non-reactive case (Fig. \ref{finger length}).
This can be explained by non-linearities and interactions between fingers that can slow down the vertical progression of their tips in the host solution.
In addition, for all measurements $t_{NL}$, $t_0$ and $\dot {z_m}$, convection starts earlier and fingers progress more slowly when $R_A=-1$ than for $R_A=1$, probably because of the buoyantly stable zone between the interface and the reaction front. 

In summary, classifying the effects of reactions on convection can be based on different criteria. 
This classification depends on whether we are interested in the onset of the convective instability, in the time when convection becomes visible or in the average progression of the fingers in the host fluid. 
For intermediate values of $\Delta R_{CB}$ between 0.1 and 0.5, convection can indeed start earlier but fingers still progress more slowly than in the non-reactive case. 
This result highlights that the criterium chosen for the classification depends on what is required for the application: earlier developing or more intense convection.

\subsection{Wavelength of pattern}\label{finger width}
\begin{figure}[htbp]\centering
	\includegraphics[width=\columnwidth]{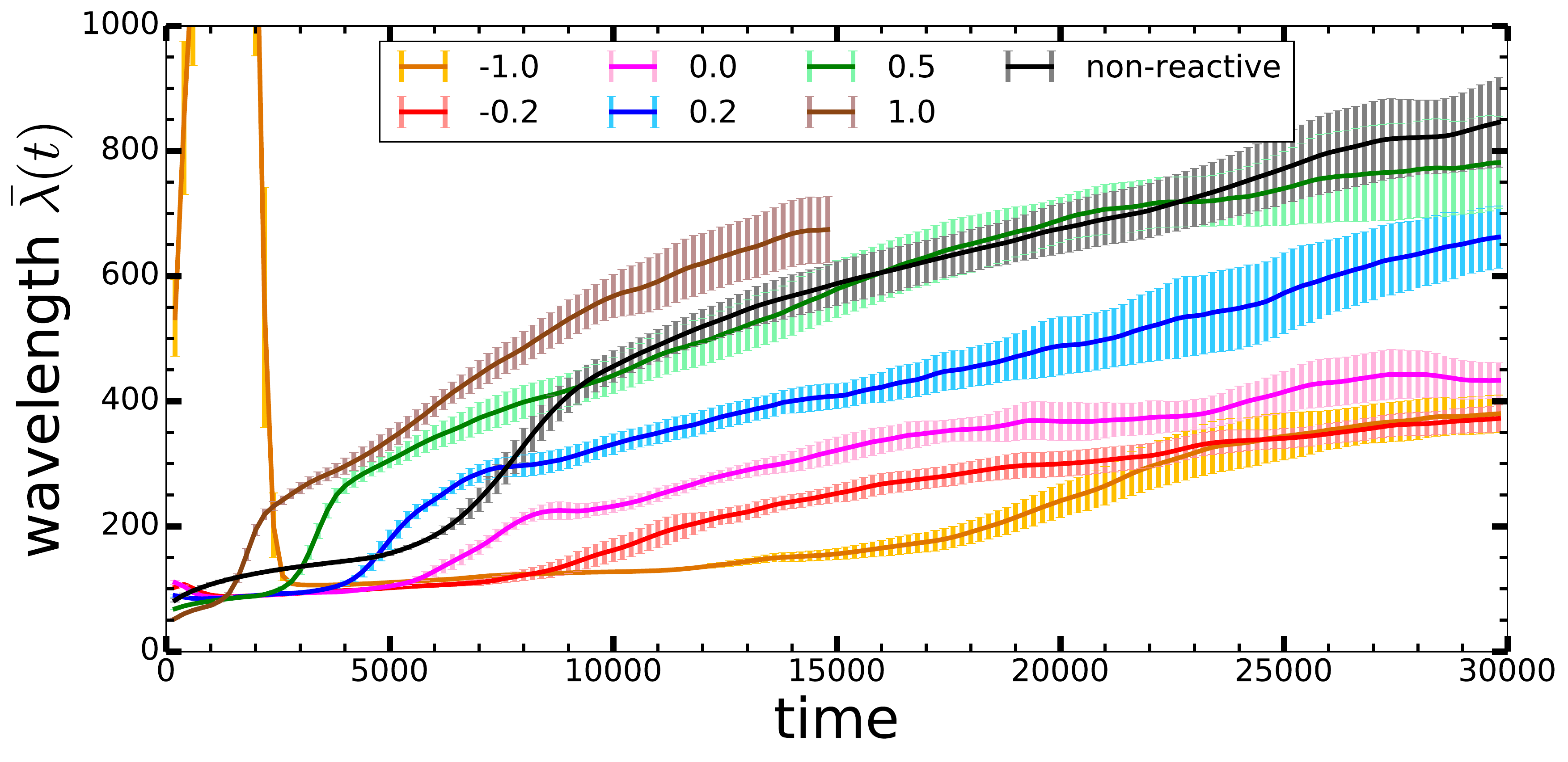}
	\caption{Power-averaged wavelength $\bar \lambda(t)$ as a function of time for $R_A=1$ and different $\Delta R_{CB}$ indicated in the graph.
	 }
	\label{fig:wavelength}
	\end{figure}

The power-averaged mean wavelength $\bar{\lambda}$ of the fingering pattern is computed as \cite{dewit04,tan86}
\begin{equation}\label{eq:meanWavelength}
\bar{\lambda}(t) = \left(\frac{\int_{1/L}^{1/(2dx)} \nu |\mathcal{F}(\bar{\rho})|^2 \diff\nu}{\int_{1/L}^{1/(2dx)} |\mathcal{F}(\bar{\rho})|^2 \diff \nu}\right)^{-1},
\end{equation}
where $\nu=1/\lambda$ is the number of wavelengths per unit distance and $\mathcal{F}(\bar{\rho})$ is the Fourier transform of the vertically averaged density profile $\bar{\rho}(x,t)$ evaluated as $\bar{\rho}(x,t) = \frac{1}{H}\int_{0}^{H} \!\rho(x,z,t) \diff z$.
In the following figures, the results are shown until the start of the shutdown regime occurring after fingers have touched the bottom of the solution. 
The wavelength $\bar{\lambda}$ typically increases over time as shown in Fig. \ref{fig:wavelength}, i.e. the number of fingers decreases as fingers become wider or merge with each other. 
This increase follows a $\sqrt{t}$ trend in the first regimes of the convective dynamics, i.e. diffusive, linear growth and flux growth, when the roots of the fingers remain mostly immobile. 
At later times corresponding to the merging regime, $\bar{\lambda}$ increases approximately linearly with time.
During this linear increase, two different slopes can be distinguished: the first one is larger, expressing intense merging, while the second one is smaller. 
For example in case NR, between $t \approx 7500$ and 10000 the wavelength increases from 200 to 450 over this time period of 2500, while after $t \approx 10000$, the wavelength increases only up to 750 over a period of 15000 (see Fig. \ref{space-time}). 
The slope of $\bar{\lambda}(t)$ increases with $\Delta R_{CB}$, meaning that fingers merge faster. 
In particular, the intense merging can occur faster (approximately when $\Delta R_{CB} \geq 0.5$) or more slowly (for $\Delta R_{CB} \leq 0.2$) than in NR case.

By comparing Figs. \ref{fig:wavelength} and \ref{finger length}, we note that the vertical and horizontal dynamics are linked. 
In the merging regime, fingers advance more slowly but merge faster. 
When two fingers merge, the resulting finger is denser as solute coming from two fingers has accumulated in only one finger, a more confined region of space. 
Therefore, during the next stage, the resulting denser fingers sink faster.  
They also merge more slowly, probably due to their amplified velocity. 
However, for $\Delta R_{CB} \leq 0.2$, fingers do not accelerate after intense merging, probably because the consequent increase of the average finger weight is not large enough to significantly affect the vertical finger velocity.

The sharp increase of the wavelength at short times before fingers are visible, as shown in Fig. \ref{fig:wavelength} for $R_A=1$ and $\Delta R_{CB}=-1$, might seem surprising but can be explained as follows.
Because perturbations are initially dampened, the largest wavelength corresponding to the width $L$ of the system has the largest Fourier amplitude, which increases its weight in the computation of the power-averaged wavelength. 
This does not happen for larger $\Delta R_{CB}$ as the time when we compute the first $\bar \lambda(t)$ is larger than the onset time of the instability.

We now compare the wavelength $\bar \lambda$ in the presence and in the absence of reactions. 
During the linear growth of the instability before merging, $\bar \lambda$ is always smaller in the reactive case than for NR case, as already highlighted in Section \ref{fingering dynamics} for specific cases. 
However, as soon as merging starts, $\bar \lambda$ can become larger than its non-reactive counterpart depending on the value of $\Delta R_{CB}$, which affects the merging rate as explained here above.

In conclusion, we have highlighted that the horizontal dynamics characterized by the mean wavelength of the fingering pattern are correlated to the vertical dynamics of the finger tips in the solution. 
In addition, although the dynamics of the pattern remain similar in all cases, changing the value of $\Delta R_{CB}$ can modify the wavelength in the linear regime and the merging rate in the successive intense and less intense merging regimes.
Our results also show that the conclusions drawn from previous studies, i.e. reactions decrease the number of fingers in the pattern \cite{budroni14,loodts15}, are not always true in the fully developed non-linear regime when fingers significantly interact with each other. 

\subsection{Reaction zone dynamics}\label{reaction zone dynamics}
Now that we have analyzed the effect of reaction on the convective fingering dynamics, we turn to the influence of convection on the dynamics of reaction fronts.
We aim to characterize how and where in the solution the dissolving species A is converted into product C.
Recently, we have investigated such reaction dynamics in the absence of convection.
To describe the evolution of these reaction-diffusion (RD) fronts, we have derived analytical expressions for the concentration profiles, valid when reaction becomes limited by diffusion, i.e. for times sufficiently larger than the chemical time scale \cite{loodts14prl, loodts15, loodts16}.
In this limit, the position $z_f$ of the reaction front evolves in time as
\begin{equation}\label{eq:front}
z_f = 2 \eta_f \sqrt{t},
\end{equation}
where $\eta_f$ is a constant depending on the control parameters of the problem. 
Here, since all species diffuse at the same rate, $\eta_f$ depends only on the ratio $\beta$ between the initial concentration $B_0$ and the solubility $A_0$ as $\eta_f  =$ erfinv$(1/(1+\beta))$, which is $\approx 0.48$ when $\beta = 1$.
The RD front delimits two zones: a ``reacted'' zone above the front, rich in dissolving species A and product C, and an ``unreacted'' zone below the front, with mostly reactant B as well as some C diffusing towards the bulk solution. 
In time, the front moves from the interface towards the bulk of the solution as reactant B is progressively depleted. 
We now examine whether the reaction front $z_f$ still progresses as $\sqrt{t}$ in the solution when convection affects the transport dynamics, and evaluate whether modifying $\Delta R_{CB}$ can alter the progression of this reaction front or the evolution of its width.

\begin{figure}[hbtp]\centering
\includegraphics[width=\columnwidth]{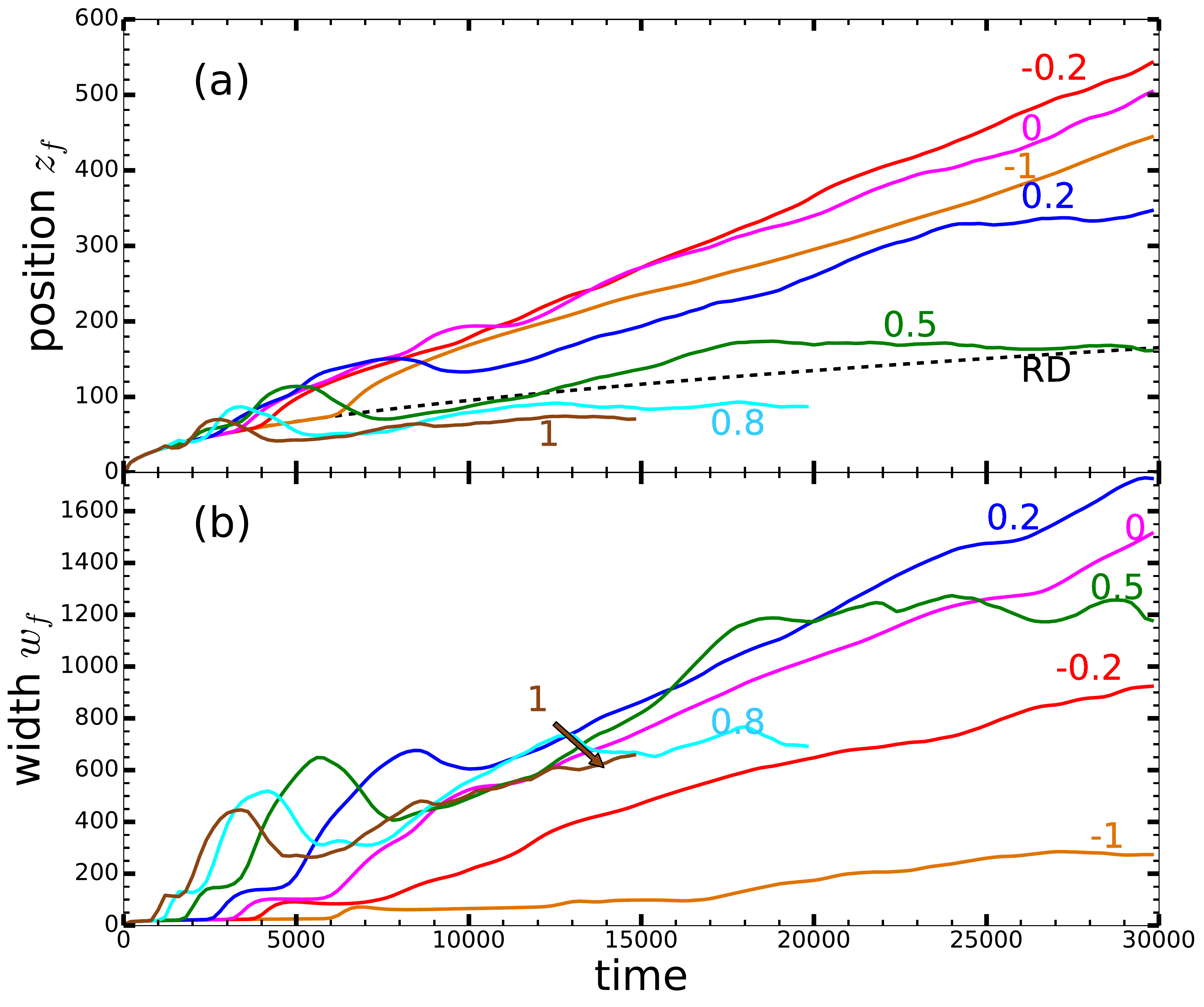}
\caption{Temporal evolution of the properties of the reaction front for $R_A=1$ and different $\Delta R_{CB}$ indicated on the graph: (a) position $z_f$ \labelcref{eq:reactionFront} with the RD counterpart \labelcref{eq:front} plotted as a dashed curve, and (b) width $w_f$ \labelcref{eq:reactionWidth}.
	}
\label{reactionFront}
\end{figure}

To quantify the dynamics of the reaction zone in the fully developed non-linear regime, we define the position $z_f$ of the reaction front as the first moment of the horizontally-averaged reaction rate profile $\bar{r}(z)$ (see Eq. \eqref{eq:rbar}): 
\begin{equation}\label{eq:reactionFront}
z_f = \frac{\int_0^{H} z \bar{r}(z) \diff z}{\int_0^{H} \bar{r}(z) \diff z}, 
\end{equation}
and the width $w_f(t)$ of the reaction front as the  width of $\bar r(z,t)$ at $\xi=0.1$ of its maximum value, proportional to the second moment of $\bar{r}(z)$ around $z_f$:
\begin{equation}\label{eq:reactionWidth}
	w_f = 2 \left(2 \ln (1/\xi) \frac{\int_0^{H} (z-z_f)^2 \bar{r}(z) \diff z}{\int_0^{H} \bar{r}(z) \diff z}\right)^{1/2}.
\end{equation}
After some time, the reaction front position deviates from the diffusive prediction \labelcref{eq:front}, as illustrated in Fig. \ref{reactionFront}a.
For $\Delta R_{CB} \leq 0.2$, including R1 case ($R_A=1$, $\Delta R_{CB}=-1$, Figs. \ref{densityField_rc-1}-\ref{zprofiles_rc-1}), the reaction front starts to evolve as $t$ and thus moves faster than the RD prediction because of convection, which is coherent with what was observed in horizontal setups where gravity currents occur \cite{rongy08, rongy10}.
For $\Delta R_{CB} > 0.5$, including R2 case ($R_A=1$, $\Delta R_{CB}=1$, Figs. \ref{densityField_rc1}-\ref{concentrationField_rc1}), the reaction front moves backwards and then stays close to the interface, because amplified convection brings fresh reactant to the interface efficiently. 
This result is coherent with the concentrations fields in Fig. \ref{concentrationField_rc1} illustrating that the dissolving species A is consumed as soon as it enters the host solution. 
Fig. \ref{reactionFront}b shows that the width $w_f$ of the reaction zone increases in time and that this increase varies non monotonically with $\Delta R_{CB}$. 
The largest values of $w_f$ are obtained for intermediate values of $\Delta R_{CB}$ (0 -- 0.5). 
The thinnest reaction zones are thus observed when $\Delta R_{CB}$ is small ($< 0$) and the reaction zone cannot extend due to the minimum of density, or when $\Delta R_{CB}$ is large ($> 0.5$) and reaction is particularly efficient due to enhanced convective transport.   

In summary, we have shown that convection does not always accelerate the progression of the reaction front in the host solution, depending on the value of $\Delta R_{CB}$. 
When $\Delta R_{CB}$ is large, the reaction takes place in a thin stationary zone close to the interface. 
When $\Delta R_{CB}$ is small, the reaction zone is also narrow but moves progressively to the bulk at a faster rate than in the absence of convection. 
The reaction zone is larger for intermediate values of $\Delta R_{CB}$. 
This means that modifying the composition of the solution, thus impacting $\Delta R_{CB}$, qualitatively affects the reaction zone dynamics during dissolution-driven convection. 

\section{Storage rate in the presence of reaction}\label{storage rate}
In many applications such as for example CO$_2$ sequestration \cite{huppert14,emami15}, it is desirable to accelerate the mixing between the dissolving phase and the host phase. 
Convection increases the mixing between both phases as it increases the movement of the dissolving species A further away from the interface, and increases the flux of A towards the host phase.
Moreover, the reaction is expected to also increase the intake of A through consumption.
In this context, what are the contributions of convection on the one hand, and of reaction on the other hand, to the evolution of the quantity of dissolved A over time?
Is it possible to affect this evolution by selecting given reactants?
Does convection coupled with reaction increase the degree of mixing compared with the non-reactive or diffusive-only cases? 
To answer those questions, we examine the coupled impact of convection and reaction on the evolution of the quantity of dissolved A and more globally on the storage rate of A into the host solution.

\subsection{Volume-averaged concentrations}
To quantify the storage rate during convective dissolution, we compute the volume-averaged concentration $\langle c_i\rangle $ of species $i$ as a function of time as 
\begin{equation}\label{eq: <ci>}
\langle c_i\rangle   = \frac{1}{V} \int_{\Omega} c_i \diff V.
\end{equation}
From this definition \labelcref{eq: <ci>}, we see that, initially, $\langle A\rangle$ and $\langle C\rangle$ are equal to zero as the dissolving species A and the product C are not present into the host solution yet, while $\langle B\rangle$ is equal to $\beta$.
When the host phase is saturated in A, $\langle A\rangle$ is equal to 1 which corresponds to the dimensionless solubility of A in the host solution and thus to its maximum possible concentration.
When the reaction is complete, all reactant B has been converted to C as A keeps dissolving into the host phase, so that $\langle C\rangle=\beta$ and $\langle B\rangle=0$.

\begin{figure}[tbhp]\centering
  \includegraphics[width=\columnwidth]{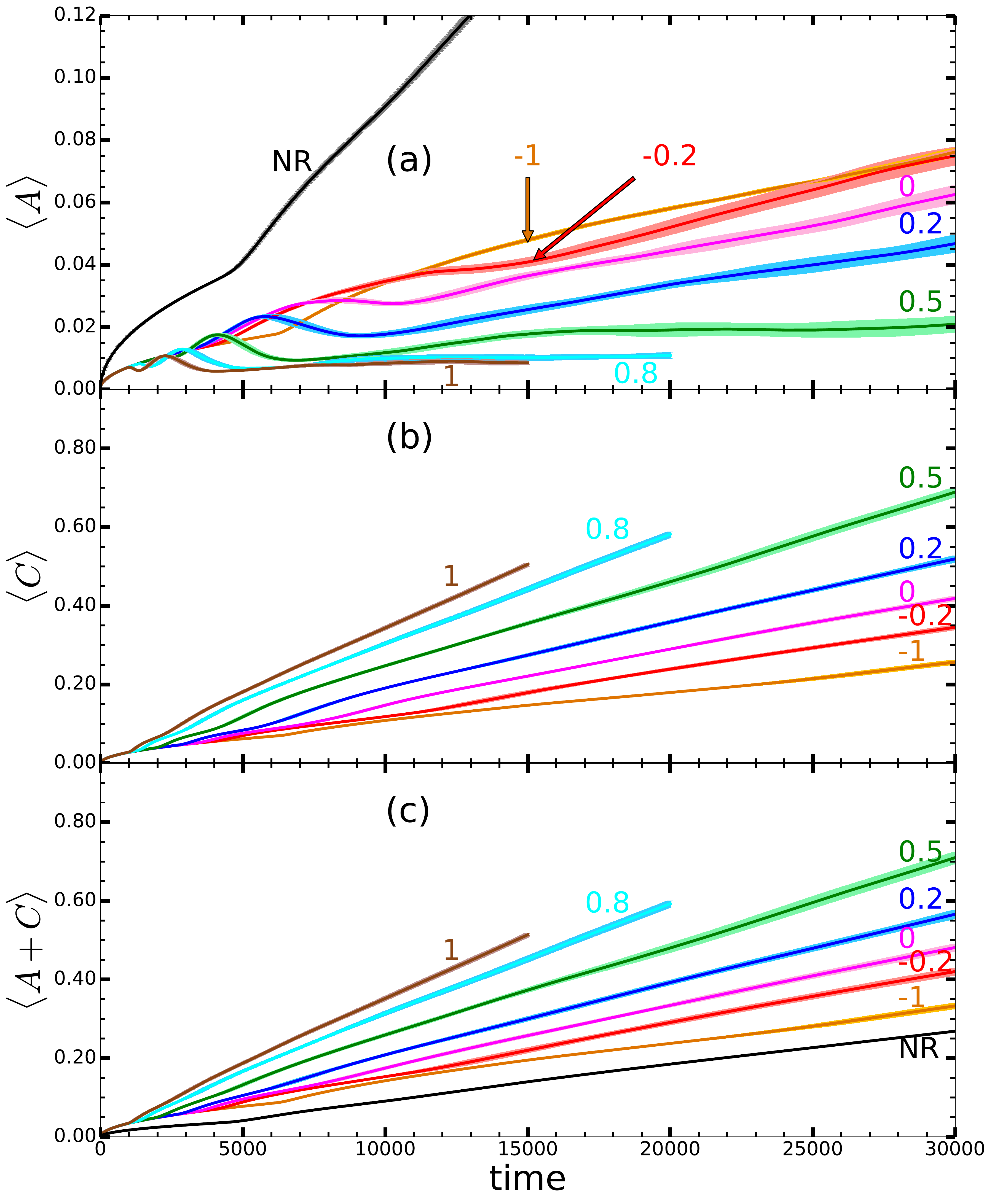}
\caption{Volume-averaged concentrations $\langle i\rangle $ as a function of time, for $R_A=1$ and different $\Delta R_{CB}$ indicated in the graph.}\label{mass}
\end{figure}

$\langle A\rangle $ increases in time as species A progressively dissolves into the solution (Fig. \ref{mass}a).
This increase is smaller in the reactive case compared to its non-reactive counterpart, because A is consumed by the reaction.
When $\Delta R_{CB}$ is amplified, the increase of $\langle A\rangle $ over time becomes slower; actually, for $\Delta R_{CB} \geq 0.5$, $\langle A\rangle $ is nearly constant for a time period of at least $\approx$ 10000.
We explain this steady-state regime in Section \ref{sec: flux}.
In addition, as C is produced by the reaction, $\langle C\rangle $ increases and this increase is faster when $\Delta R_{CB}$ increases (Fig. \ref{mass}b).
We define the amount of stored A as $\langle A+C\rangle $, which reflects that species A can be stored in the form of dissolved A or product C.
$\langle A+C\rangle $ increases more slowly in the non-reactive case than in the reactive case (Fig. \ref{mass}c), 
which means that chemical reactions improve the efficiency of the phase transfer by accelerating the storage process.
Most of the stored A is in the form of product C and there is only few dissolved A, as  concluded from the comparison of Fig. \ref{mass}a-c.
Further, the larger $\Delta R_{CB}$, the larger is the quantity of A stored as product C.

To better understand why $\langle c_i\rangle $ evolve as shown in Fig. \ref{mass}, we derive evolution equations for these averaged values by integrating over the whole spatial domain the equations \eqref{eq: a1}-\eqref{eq: c1} of evolution for the solute concentrations and taking into account the boundary conditions \eqref{cb u}-\eqref{cb c}.
In the non-reactive case, the quantity of dissolved A in the host phase evolves in time as \cite{hidalgo12}:
	\begin{equation}\label{eq: <A> _time nr}
	\frac{\partial \langle A\rangle }{\partial t} = \frac{J}{H},
	\end{equation}
which expresses that $\langle A\rangle$ increases due to the dissolution flux $J$ scaled by the depth of the system $H$, as a deeper system takes more time to achieve saturation.
In the presence of reactions, the $\langle c_i\rangle $ evolve as
\begin{eqnarray}\label{eq:<i>}	
	 \frac{\partial \langle A\rangle }{\partial t} &= \frac{J}{H} - r,  \label{eq:<A>} \\
	\frac{\partial \langle C\rangle }{\partial t} &= r,  \label{eq:<C>} \\
	\frac{\partial \langle A+C\rangle }{\partial t} &= \frac{J}{H},\label{eq:storage}
\end{eqnarray}
where $r$ is the volume-averaged reaction rate $\langle AB\rangle $.
Equation \eqref{eq:<i>} express that the evolution of the quantities of the solutes depends on the dissolution flux $J$ and the global reaction rate $r$, which we analyze here below.

\subsection{Dissolution flux and volume-averaged reaction rate}\label{sec: flux}
A dissolution flux through an interface can consist of two different contributions: the convective flux $-A \bm{u} \cdot \bm{n}$, where $\bm{n}$ is the unit vector perpendicular to the interface and oriented towards the outside of the host phase; and the diffusive flux $-\frac{\partial A}{\partial z}$ across the interface.
The convective flux across the interface is zero by definition in the theoretical modeling of a monophasic system (see boundary condition \labelcref{cb u}) and negligeable in biphasic systems \cite{rongy12}. 
We thus compute the interface-averaged dissolution flux $J$ of species A as
	\begin{equation}\label{eq: flux}
	J  = -\frac{1}{L} \int_{0}^{L} \left.\frac{\partial A}{\partial z}\right\rvert_{z=0} \diff x.
	\end{equation}

We evaluate the diffusive dissolution flux $J_D$ in the non-reactive case as \cite{loodts16} 
\begin{equation}\label{eq: fluxNr}
J_{D}  = \frac{1}{\sqrt{\pi t}},
\end{equation}
 and the reaction-diffusion flux $J_{RD}$ as \cite{loodts16} 
\begin{equation}\label{eq: fluxR}
J_{RD}  = \frac{1+\beta}{\sqrt{\pi t}},
\end{equation}
here equal to $2/\sqrt{\pi t}$ as $\beta=1$. 
Equations \eqref{eq: fluxNr} and \eqref{eq: fluxR} show that for all time $t$, there is a constant ratio $J_{RD}/J_D = 1 + \beta$ between the reactive $J_{RD}$ and non-reactive $J_D$ fluxes. 
Even without convection, the reactive flux is always larger than its non-reactive counterpart as $\beta > 0$, and this difference becomes larger when $\beta$ is amplified. 
Chemical reactions amplify the flux of A across the interface because the consumption of A by the reaction increases the concentration gradient at the origin of the diffusive flux.

\begin{figure}[htbp]\centering
	\includegraphics[width=\columnwidth]{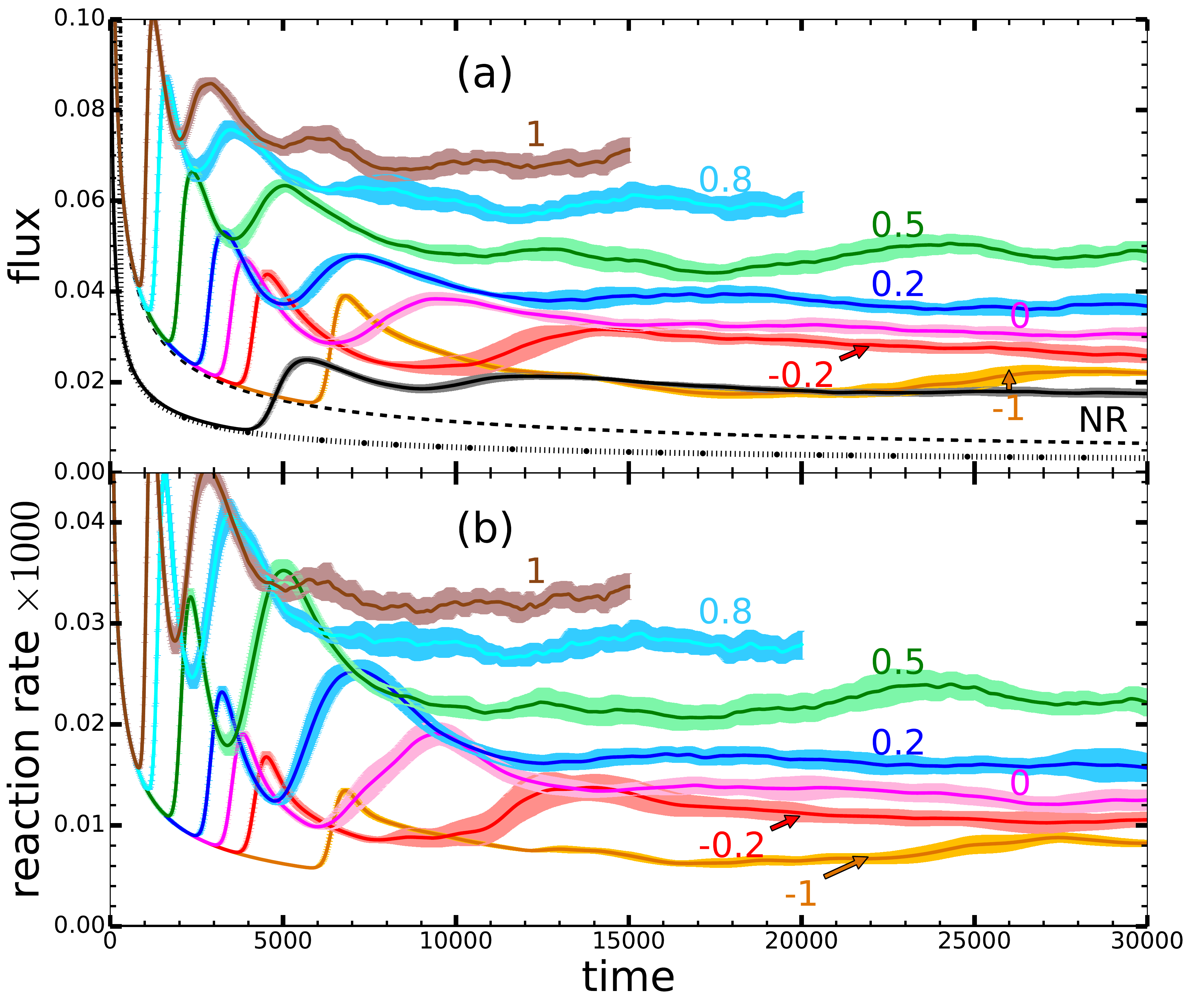}
	\caption{(a) Flux (\ref{eq: flux}) with the dotted and dashed curve representing the diffusive flux $J_D$ \labelcref{eq: fluxNr} and the reaction-diffusion flux $J_{RD}$ (\ref{eq: fluxR}), respectively; (b) reaction rate $r=\langle AB\rangle $ as a function of time for $R_A=1$ and different $\Delta R_{CB}$ indicated in the graph.
	}
	\label{flux}
	\end{figure}

As shown in Fig. \ref{flux}a, the flux initially decreases in time following Eq. \eqref{eq: fluxNr} (dotted curve) or Eq. \eqref{eq: fluxR} (dashed curve) as long as diffusion remains the dominant transport process. 
After some time, the flux starts to increase because of convection and eventually fluctuates around a steady-state value $J^*$.
The temporal evolution of $J$ is qualitatively the same for non-reactive and reactive cases.
We compute the steady-state flux $J^*$ as the average over the last time interval $> 3000$ when the variation of the flux with time (least-squares fitted slope) is no more than a small threshold, here arbitrarily chosen as $10^{-6}$. 
Some fluctuations occur around $J^*$ due to reinitiation: the flux increases when protoplumes form as the boundary layer then becomes thinner, and conversely the flux decreases when protoplumes merge with older fingers \cite{slim14}.  
We compute $J^*$ in the non-reactive case as 0.019, in agreement with previous studies \cite{slim14, pau10,elenius12}.

The evolution of the volume-averaged reaction rate $r=\langle AB\rangle $, shown in Fig. \ref{flux}b, is similar to that of the dissolution flux $J$ (Fig. \ref{flux}a): when $J$ increases, $r$ increases too.
$J$ and $r$ thus appear to be correlated as a more efficient reaction increases the concentration gradient at the origin of the dissolution flux.
After a while $r$ also fluctuates around a steady-state value $r^*$, which we compute similarly to $J^*$ but with a smaller threshold ($4 \times 10^{-10}$) for the slope of $r(t)$ given that $r$ is three orders of magnitude smaller than $J$. 

\begin{figure}[htbp]
\centering
\includegraphics[width=\columnwidth]{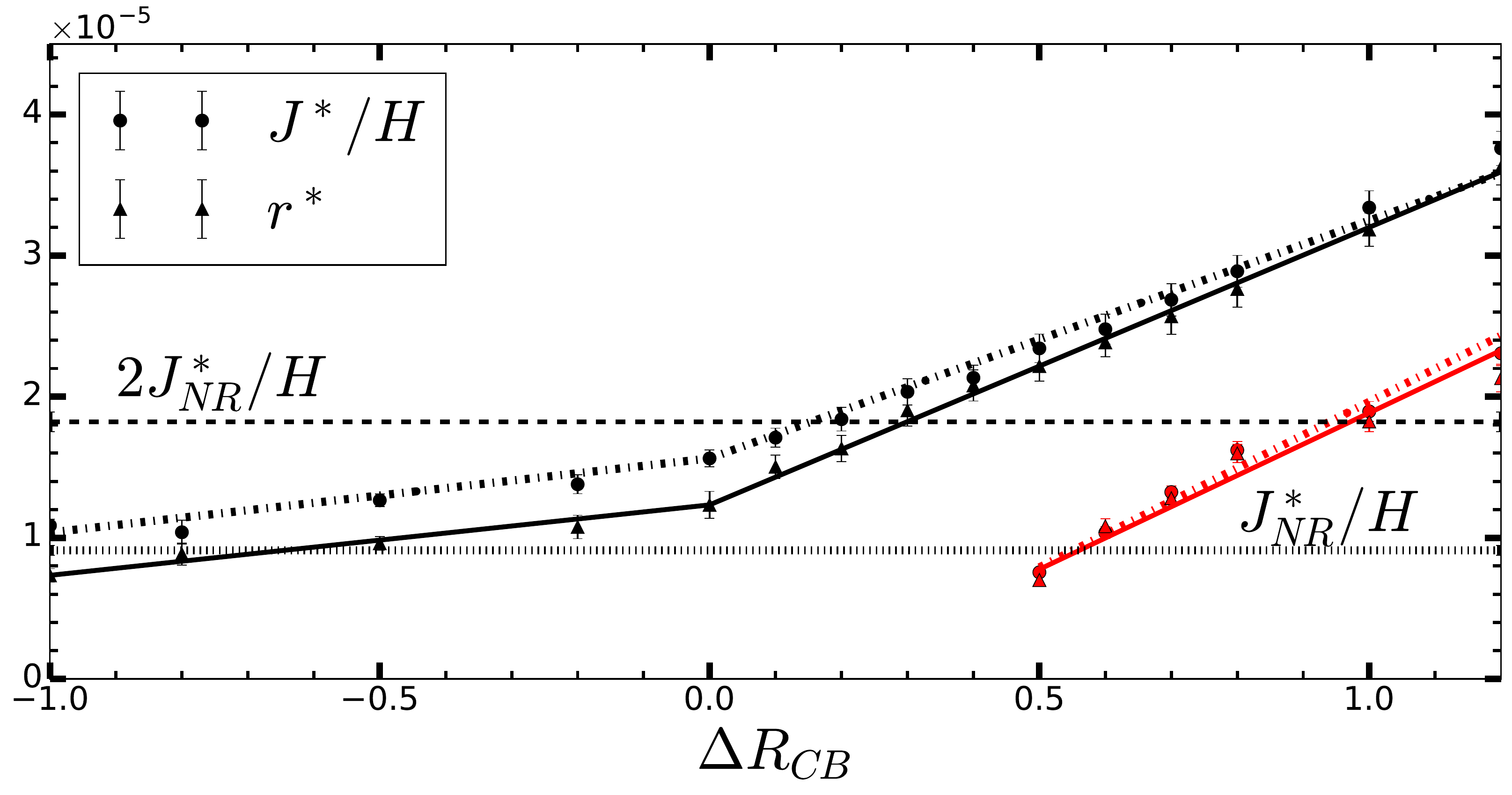}
\caption{Comparison between $J^*/H$, steady-state flux scaled with regard to the height of the host solution (dashed-dotted curves) and the steady-state reaction rate $r^*$ (full curves) for $R_A=1$ (black), $R_A=-1$ (red) and $\beta =1$.
The steady-state flux $J^*_{NR}$ in the non-reactive case is plotted as a dotted line.
}
\label{constantReactionRate}
\end{figure}

We now analyze the variation of the steady-state reaction rate $r^*$ (full curve) and scaled flux $J^*/H$ (dot-dashed curve) as a function of $\Delta R_{CB}$ in Fig. \ref{constantReactionRate}. 
Both $J^*$ and $r^*$ increase with $\Delta R_{CB}$, with a change of slope at $\Delta R_{CB} =0$. 
When $R_A=1$, this increase can be described by the empirical fits between $\Delta R_{CB} =-1$ and $1.2$ (see lines in Fig. \ref{constantReactionRate}):
	\begin{eqnarray}\label{eq: J*r* ra1}	
	 \Delta R_{CB} \leq 0: &\quad r^* = 1.2 \times 10^{-5} + 0.5\times 10^{-5} \Delta R_{CB},\label{r* lina} \\
	 &\quad J^*/H = 1.6 \times 10^{-5} + 0.5\times 10^{-5} \Delta R_{CB}; \\
	 \Delta R_{CB} \geq 0: &\quad r^* = 1.2 \times 10^{-5} + 2.0\times 10^{-5} \Delta R_{CB}, \label{r* linb}\\
	 &\quad J^*/H = 1.6 \times 10^{-5} + 1.7\times 10^{-5} \Delta R_{CB} \label{J* linb};
\end{eqnarray}
and similarly when $R_A=-1$ between $\Delta R_{CB} =0.5$ and $1.2$ (see lines in Fig. \ref{constantReactionRate}):
	\begin{eqnarray}\label{eq: J*r* ra-1} 	
	 r^* & = -0.3 \times 10^{-5} + 2.2 \times 10^{-5} \Delta R_{CB}, \label{r* linc}\\
	 J^*/H & = -0.4\times 10^{-5} + 2.3 \times 10^{-5}\Delta R_{CB} \label{J* linc}.
	\end{eqnarray}
The increase of $J^*$ and $r^*$ with $\Delta R_{CB}$, described by Fig. \ref{eq: J*r* ra1,eq: J*r* ra-1}, can be explained as follows.
When the contribution of the product C to the density increases, convection starts earlier and reaches a larger amplitude (Fig. \ref{u2}). 
This increased convection accelerates the transport of fresh reactant B to the interface, which increases the efficiency of the reaction and thus the dissolution flux.

For any value of $\Delta R_{CB}$, the steady-state reaction rate $r^*$ is smaller or equal to the scaled steady-state dissolution flux $J^*/H$: the reaction is limited by the dissolution rate of A into the host solution.  
When $R_A=1$, we note that if $\Delta R_{CB} < 0.5$, $r^*$ is typically smaller than $J^*/H$, so that A accumulates in the host fluid, and $\langle A \rangle$ increases in time (Eq. \eqref{eq:<A>}, Fig. \ref{mass}a).
However, as $r^*$ increases faster than $J^*/H$ when $\Delta R_{CB}$ increases (see Eqs. \eqref{r* linb}-\eqref{J* linb}), we see that above $\Delta R_{CB} \geq 0.5$, $r^* \approx J^*/H$ so that $\langle A\rangle$ remains constant in the steady-state flux regime (Eq. \eqref{eq:<A>}, Fig. \ref{mass}a): as soon as A enters the host fluid, it is consumed by the reaction with B.
When $R_A=-1$, $J^*/H \approx r^*$ for all cases as shown in Fig. \ref{constantReactionRate} and expressed by Eqs. \eqref{r* linc}-\eqref{J* linc}. 

We can explain the change of slope appearing around $\Delta R_{CB}=0$ in $r^*(\Delta R_{CB})$ and $J^*(\Delta R_{CB})$ for $R_A=1$ (see Eqs. \eqref{r* lina}-\eqref{J* linb}) as follows.
Let us first recall that when all species diffuse at the same rate, the density at the interface is given by $R_A +\Delta R_{CB} \beta$, the density at the reaction front is $\Delta R_{CB} \beta$ and the density of the bulk solution is 0 (see Ref. \citenum{loodts15} for more details). 
For $\Delta R_{CB} < 0$, the RD density profile in the host solution has a minimum of density. 
The difference of density at the origin of the instability is then equal to $R_A$ and corresponds to the one between the interface and the minimum at the reaction front. 
Increasing $\Delta R_{CB}$ does not modify this difference but decreases the amplitude of the stabilizing barrier, i.e. the difference of density $-\Delta R_{CB}$ between the minimum and the bulk solution. 
By contrast, for $\Delta R_{CB} \geq 0$ the density profile in the solution is monotonic. 
The density difference at the origin of the instability is then between the interface and the bulk solution, and is equal to $\Delta R_{CB} \beta$.
This explains why increasing $\Delta R_{CB}$ affects $J^*/H$ and $r^*$ less when $\Delta R_{CB} < 0$ and more when $\Delta R_{CB} \geq 0$.

To analyze the effects of reaction on the steady-state scaled flux $J^*/H$, we compare $J^*/H$ to its non-reactive counterpart $J^*_{NR}/H$ for $R_A=1$, plotted in Fig. \ref{constantReactionRate} as a dotted line. 
For all $\Delta R_{CB}$, $J^*/H$ is larger in reactive cases than in NR case.
This is not surprising as already without convection, the reaction-diffusion (RD) flux (Eq. \eqref{eq: fluxR}) is always $(1+\beta)$ times larger than the diffusive flux (Eq. \eqref{eq: fluxNr})\cite{loodts16}.
To understand whether this increased steady-state flux is due to RD effects only or also due to convection amplified by reaction, we compare $J^*/H$ to the theoretical value of $(1+\beta) J^*_{NR}/H$ for the reactive case where only RD effects would affect the dissolution flux.
As $J^*/H$ increases with $\Delta R_{CB}$, it becomes larger than $(1+\beta) J^*_{NR}/H$ for a critical value of $\Delta R_{CB}=0.2$, which corresponds to the critical value computed from the onset time $t_0$ (see Fig. \ref{finger velocity}). 
In other words, below that critical value, the increase of $J^*$ in the reactive case is due to RD effects only, while above that critical value, the increase of $J^*$ is also due to convection amplified by reaction.

In summary, the presence of a reactant B in the solution always accelerates the storage of A, even before dissolution-driven convection develops.
This can completely change any long-term predictions concerning the fate of A into the host phase. 
However, depending on the composition of the host solution (impacting $\Delta R_{CB}$), the convection can be amplified by the reaction so that the storage occurs even faster than predicted on the basis of reaction and diffusion alone. 

\section{Conclusion}\label{NLconclusion}
We have numerically characterized how the interplay between an A+B$\rightarrow$C reaction and dissolution-driven convection affects the spatio-temporal non-linear dynamics of fingering and reactions fronts as well as the storage efficiency of a phase A into a host fluid phase.
Compared to the non-reactive case, reactions can accelerate or slow down fingering development depending on $\Delta R_{CB}$ quantifying the difference between the solutal contributions of product C and of reactant B to density. 
We have revisited the classification previously established on the basis of a linear stability analysis \cite{loodts14prl,loodts15}, showing that in some reactive cases convection starts earlier but fingers progress more slowly in the solution than in the non-reactive case, due to non-linear interactions between fingers at later times.

In addition, we have presented different types of convective dynamics, depending on the type of RD density profile building up in the host phase before convection sets in. 
We have highlighted that when the product C contributes less to the density than the reactant B, a local minimum of density constrains fingering.
Hence the reaction front follows the finger tips, but moves downwards at a faster pace than without convection. 
The fingering pattern originates then mainly from the dissolving species A, progressively accumulating in the host solution as its dissolution takes place at a larger pace than its consumption by the reaction.
By contrast, when C contributes much more to the density than B, the reaction, which has a destabilizing effect, takes place mostly in a stationary zone close to the interface.
In the steady-state regime, the dissolving species A is consumed as soon as it enters the host solution such that the quantity of A in the host phase remains constant and the fingering pattern is formed mostly by the denser product C.
We have also discussed the case where the reaction is at the origin of dissolution-driven convection, because a local maximum forms in the density profile. 
Due to this specific density profile, fingers then have a characteristic shape and form at a given distance below the interface with phase A. 

For all reactive cases, the steady-state flux is larger than its non-reactive counterpart because the consumption of the dissolving species A by the reaction amplifies the concentration gradient at the origin of the dissolution flux. 
Both steady-state flux and reaction rate increase with $\Delta R_{CB}$ as convection develops earlier and becomes more intense, amplifying the mixing between the phase A and the host phase.  
Although this model could be extended to include differential diffusivity effects \cite{loodts16} and a variable solubility depending on the solute concentrations \cite{loodts14chaos}, our results already highlight that selecting an appropriate composition for the host phase allows to maximise the positive effect of reaction for amplifying the storage rate of A into the host fluid.

These conclusions are useful to predict the fate of CO$_2$ during its sequestration in subsurface formations and to select storage sites with geochemical reactions optimal in enhancing convective dissolution.
Knowing the kinetic properties of the geochemical reaction is not enough to quantify the storage rate; one also needs to know the contributions to density of all dissolved species. 
For other applications where convection enhances mass transfer, controlling the properties of the dissolution-driven convection should become possible by selecting the appropriate reactant to be dissolved in the host solution.





\section*{Acknowledgments}
Funding by PRODEX, ARC CONVINCE, ARC PIONEER, MIS-FNRS PYRAMID and PDR-FNRS FORECAST projects is gratefully acknowledged. 
We thank V. Moureau and G. Lartigue for their training and support concerning YALES2 code, as well as F. Brau, C. Rana and V. Upadhyay for useful comments and scientific discussions on this study.  


\providecommand*{\mcitethebibliography}{\thebibliography}
\csname @ifundefined\endcsname{endmcitethebibliography}
{\let\endmcitethebibliography\endthebibliography}{}

\end{document}